\documentclass[aps, pra, 
			   final,  
			   twocolumn]{revtex4-1} 

\usepackage{amsmath}
\usepackage{amsfonts}
\usepackage{amssymb}
\usepackage{gensymb}
\usepackage{graphicx}
\usepackage[utf8]{inputenc}
\usepackage[colorlinks=true, linkcolor=black, urlcolor=blue, citecolor=black, anchorcolor=black]{hyperref}
\usepackage{xcolor}

\usepackage{enumitem}
\setlist{nosep}

\begin{document}
\title{Pump-triple sum-frequency-probe spectroscopy of transition metal dichalcogenides}
\author{Darien J. Morrow}
\author{Daniel D. Kohler}
\author{Yuzhou Zhao}
\author{Song Jin}
\author{John C. Wright}
	\email{wright@chem.wisc.edu}
\affiliation{Department of Chemistry,
	University of Wisconsin--Madison,
	1101 University Ave, 
	Madison, WI 53706, United States}
\keywords{THG, TSF, CMDS, MoS\textsubscript{2}}

\date{\today}

\begin{abstract}
	Triple sum-frequency (TSF) spectroscopy measures multidimensional spectra by resonantly exciting multiple quantum coherences of vibrational and electronic states.
	In this work we demonstrate pump-TSF-probe spectroscopy in which a pump excites a sample and some time later three additional electric fields generate a probe field which is measured. 
	We demonstrate pump-TSF-probe spectroscopy on polycrystalline, smooth, thin films and spiral nanostructures of both MoS\textsubscript{2} and WS\textsubscript{2}.
	The pump-TSF-probe spectra are qualitatively similar to the more conventional transient-reflectance spectra.
	While transient-reflectance sensitivity suffers under low surface coverage, pump-TSF-probe sensitivity is independent of the sample coverage and nanostructure morphologies.
	Our results demonstrate that pump-TSF-probe is a valuable methodology for studying microscopic material systems.
\end{abstract}

\maketitle

\section{Introduction}\label{S:Introduction}

Pump-probe spectroscopy is a ubiquitous methodology for investigating the dynamics and energetics of excited systems on sub-picosecond time scales.
In a pump-probe experiment, a pump excites the system of interest and a probe interrogates the evolved system at a later time, $T$.
The differences in the probe signal with and without the pump inform on system evolution.
Most merits of a pump-probe experiment, such as sensitivity and selectivity, are determined by the choice of a specific probe methodology, of which there are many.\cite{Ulbricht_Bonn_2011, Xiong_Zanni_2009, Dietze_Mathies_2016, Bragg_Magnanelli_2016, Ceballos_Zhao_2017, Mandal_Wasielewski_2019, Sie_Lindenberg_2019, Liu_Zhu_2019,  Langer_Huber_2018,  Wang_DiMauro_2017} 
The development of Coherent Multidimensional Spectroscopy (CMDS) offers promising possibilities for new probes because CMDS methods can have increased selectivity compared to conventional methods.\cite{Wright_2011, Wright_2017, Chen_2016, Smallwood_Cundiff_2018, Cundiff_Mukamel_2013, Cho_2008, Cho_2019}
CMDS uses multiple optical interactions to create a multiple quantum coherence within the material whose optical emission is measured.
The ability/requirement to couple multiple quantum states together leads to the selectivity inherent within CMDS.
By preceding a CMDS pulse sequence by a pump, the selectivity of CMDS can be leveraged as a probe in a ``pump-CMDS-probe'' measurement.\cite{Bredenbeck_Hamm_2003, Xiong_Zanni_2009, Dietze_Mathies_2016, Mandal_Wasielewski_2019, Abraham_Gundlach_2019} 
In this paper we introduce triple sum-frequency (TSF) spectroscopy as a new probe for material systems by measuring the pump-induced TSF response of model semiconductor systems: transition metal dichalcogenides (TMDCs). 

TSF spectroscopy uses three tunable electric fields, $E_1$, $E_2$, and $E_3$ to create coherences at increasingly higher energies.
These coherences cooperatively emit a new electric field with frequency $\omega_\text{out} = \omega_1 + \omega_2 + \omega_3$ in a direction defined by phase-matching.
Scanning the multiple driving laser frequencies enables collection of a multidimensional spectrum whose cross-peaks identify dipole coupling among probed states.
The selectivity of TSF is due to the increase in output intensity achieved when the driving fields are resonant with multiple states; the multiple resonance conditions act as a ``fingerprint''.\cite{NeffMallon_Wright_2017} 
TSF has been used to investigate vibrational and electronic coupling in molecules,\cite{Handali_Wright_2018_B, Boyle_Wright_2013, Boyle_Wright_2013_001, Boyle_Wright_2014, Grechko_Bonn_2018_B, Bonn_Cho_2001}
and recently, TSF has revealed the electronic states of MoS\textsubscript{2} and the mixed-vibrational-electronic coupling of organic-inorganic perovskites.\cite{Morrow_Wright_2018, Grechko_Bonn_2018}

\begin{figure}[!htbp]
	\centering
	\includegraphics[width=\linewidth]{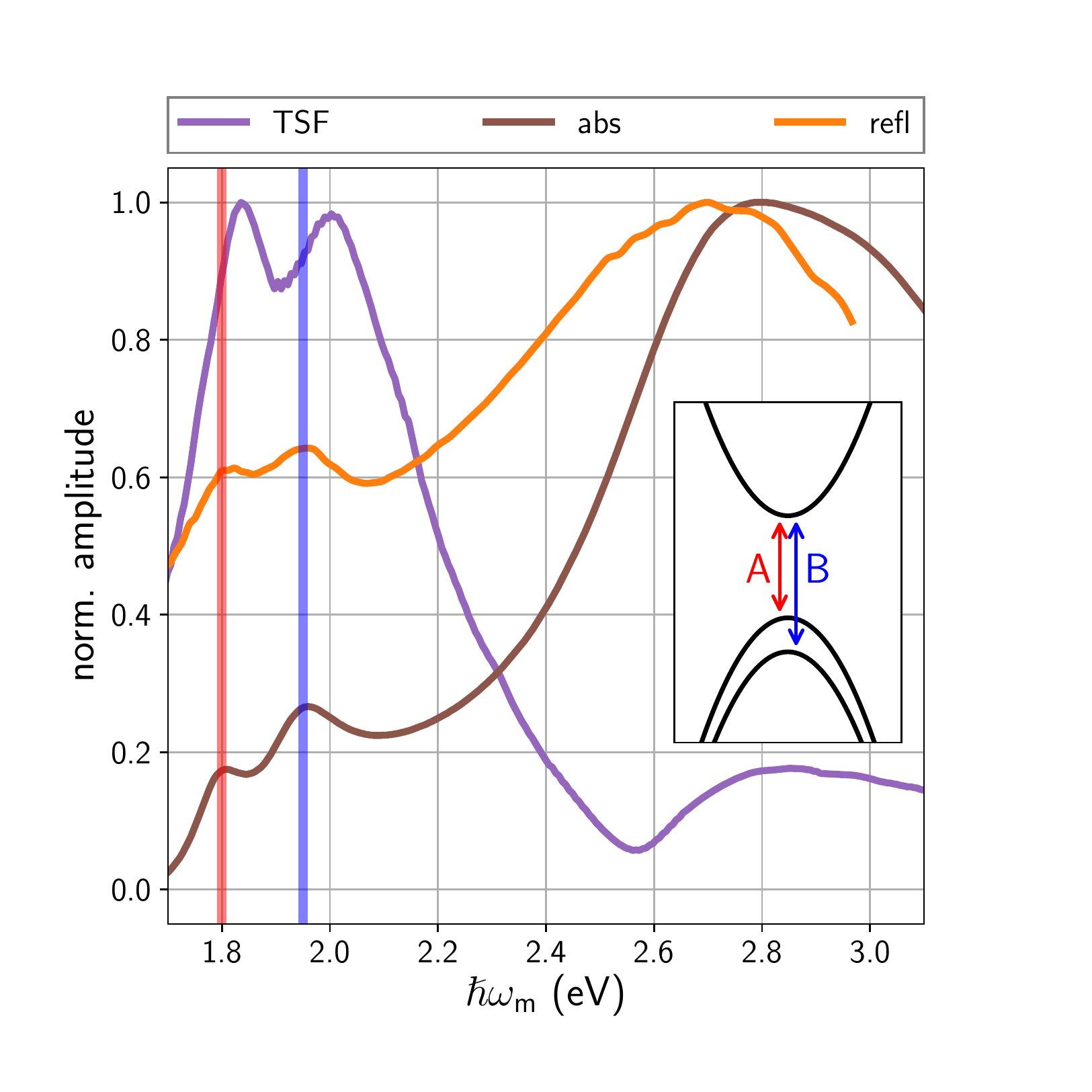}
	\caption{
		Normalized amplitude 1D spectra of MoS\textsubscript{2} thin films. 
		The absorption measurement was originally shown in \textcite{Czech_Wright_2015}.
		The TSF and reflection contrast measurements were originally shown in \textcite{Morrow_Wright_2018}.
		Vertical bars are guides to the eyes set at 1.80 and 1.95 eV. 
		The inset is a cartoon of the band structure of MoS\textsubscript{2} at the $K$ point. Only the valence bands are shown as spin-orbit-split because the splitting of the conduction bands is generally too small to be observed for MoS\textsubscript{2}.
	}\label{fig:1D}
\end{figure}

In this paper, we measure the pump-TSF-probe response of MoS\textsubscript{2} and WS\textsubscript{2}, which are layered semiconductors in the TMDC family.\cite{Mak_Heinz_2010}
The bandedge optical spectrum of MoS\textsubscript{2} is dominated by two features labeled A ($\hbar\omega_\text{A} \approx 1.8 \text{ eV}$) and B ($\hbar\omega_\text{B} \approx 1.95 \text{ eV}$) which originate from high binding energy excitonic transitions between spin-orbit split bands (see absorption spectrum and inset diagram in \autoref{fig:1D}).\cite{Wang_Urbaszek_2018, MolinaSanchez_Wirtz_2013, Qiu_Louie_2013, He_Shan_2014, Saigal_Ghosh_2016, Kopaczek_Kudrawiec_2016}
Likewise, the optical response of WS\textsubscript{2} is dominated near the bandedge by the A feature ($\hbar\omega_\text{A} \approx 2 \text{ eV}$).
The present work expands upon our previous work on the unpumped TSF response of MoS\textsubscript{2},\cite{Morrow_Wright_2018} the extensive body of harmonic generation work on TMDCs (c.f. the review by \textcite{Autere_Sun_2018} and references therein), and the innovative pump-second-harmonic-generation probe work accomplished on semiconductors.\cite{Chang_Tom_1997, Guo_Taylor_2001, McClelland_Borguet_2004, Hsieh_Gedik_2011, Tisdale_Zhu_2010, Park_Zhu_2013, Nelson_Zhu_2014, Mannebach_Lindenberg_2014}

In our previous work on the unpumped TSF response of MoS\textsubscript{2} we noted important differences between the non-linear TSF probe and conventional linear probes, such as absorption or reflection.\cite{Morrow_Wright_2018}
The intensity of homodyne detected TSF has transition dipole scaling of $\mu^8$ and state density scaling of $J^2$.
This scaling is in contrast to self-heterodyne detected absorption and reflection measurements which scale as $\mu^2$ and $J$.
The steep scaling of TSF with transition dipole compared to state density can depress substrate effects which dominate reflection measurements.
For instance, we have found that in the case of large transition dipole excitonic transitions, TSF only measures photons which originated in a single nanostructure whereas reflection measurements sense reflections from both the nanostructure and reflections from the substrate. 
Likewise, the dipole scaling of other CMDS techniques has enabled the measurement of protein structure against large backgrounds when conventional absorption measurements fail.\cite{Lomont_Zanni_2017, Alperstein_Zanni_2019}
The ability of TSF to selectively interact with large dipole transitions is highlighted in \autoref{fig:1D} for the example of MoS\textsubscript{2}. 
The absorption and reflection spectra of the  MoS\textsubscript{2} thin film are dominated by higher energy transitions with large joint density of states and low transition moments.
Conversely, the TSF spectrum (in this case $\omega_1 = \omega_2 = \omega_3 = \omega_{\text{out}}/3$, a third harmonic generation, THG, spectra) is dominated by the large transition dipole A and B excitonic transitions.

The structure of the rest of this paper is as follows:
In the Theory section we describe how to calculate transient-TSF response and graphically compare it to the response from other common spectroscopies.
In the Experimental section we describe our spectrometer and our samples.
In the Results section we present our transient-TSF measurements on TMDCs. 
We first examine how the multidimensional TSF spectrum is affected by an optical pump.
We find that the multidimensional TSF spectrum can be fully described by the one-dimensional pump-THG-probe spectrum.
We then compare pump-THG-probe to pump-reflectance-probe spectroscopy; we demonstrate that although their lineshapes appear slightly different, the same pump-induced physics can explain both spectra.
Finally we demonstrate the utility of transient-TSF for measuring TMDC nanostructures.
We finish the paper by discussing how transient-TSF might be used in the future on other systems.
 
\section{Theory}\label{S:Theory}

\subsection{The linear and non-linear probe}

In this section we present the correspondence between the reflectance and TSF of a material.
We investigate the phenomenological, microscopic properties that are responsible for the susceptibility and also how the susceptibility dictates the electric field output.
Readers interested in first-principle calculations of TMDC nonlinear suceptibility should consult refs.\cite{Taghizadeh_Pedersen_2019,Taghizadeh_Pedersen_2018,Taghizadeh_Pedersen_2017, Pedersen_2015, Soh_Mabuchi_2018}.
Our analysis uses standard perturbation theory.\cite{Boyd_2008, Bloembergen_Shen_1964}
The material polarization, $P$, is expanded in orders of the electric field, $E$:
\begin{equation}
P = \epsilon_0\left( \chi^{(1)}E + \chi^{(2)}E^2 + \chi^{(3)}E^3 + \cdots \right),
\end{equation} 
where $\chi^{(n)}$ is the $n^{th}$-order susceptibility and $\epsilon_0$ is the permittivity of free space.
The linear susceptibility, $\chi^{(1)}$, determines the response of linear spectroscopies such as absorption and reflection.
The third-order susceptibility, $\chi^{(3)}$, determines the response of non-linear spectroscopies such as TSF.

Within the dipole approximation, $\chi^{(1)}$ is constructed from a sum over all initial and final states:
\begin{equation} 
	\chi^{\left(1\right)} \left( \omega_1 \right) = 
	\sum_{a, g}
	\frac{\mu_{ag}^2}{
		\Delta_{ag}^{1}}, \label{E:chi1}
\end{equation}
where $\Delta_{ag}^{1}\equiv \omega_{ag} - \omega_1 -i\Gamma$, $\mu_{ag}$ and $\omega_{ag}$ are the transition dipole and frequency difference between states $a$ and $g$, $\Gamma$ is a damping rate which accounts for the finite width of the optical transitions, and $\omega_1$ is the driving frequency. 
We see from \autoref{E:chi1} that when the driving field is resonant ($\omega_1 = \omega_{ag}$), $\chi^{(1)}$ is large and the interaction with light is strong.

Like \autoref{E:chi1}, the TSF susceptibility is a sum over states, but we must consider three sequential excitations $g \rightarrow a \rightarrow b \rightarrow c$:
\begin{align}
	\chi^{\left(3\right)} \left(-\omega_{321}, \omega_1, \omega_2, \omega_3\right) &= 
		\mathcal{P} \sum_{c, b, a, g}
		\frac{\mu_{gc} \mu_{cb} \mu_{ba} \mu_{ag}}{
			\Delta_{gc}^{123}
			\Delta_{gb}^{12}
			\Delta_{ga}^{1}
		}, \label{E:chi3} \\
	\Delta_{ga}^{1} &\equiv \omega_{ag} - \omega_1 -i\Gamma, \nonumber \\
	\Delta_{gb}^{12} &\equiv \omega_{bg} -\omega_{21} -i\Gamma, \nonumber \\
	\Delta_{gc}^{123} &\equiv \omega_{cg}-\omega_{321} -i\Gamma,  \nonumber \\  
	\omega_{21} &\equiv \omega_2 + \omega_1 , \nonumber \\  
	\omega_{321} &\equiv \omega_3 + \omega_2 + \omega_1, \nonumber
\end{align}
where $\mathcal{P}$ is a permutation operator which accounts for all combinations of field-matter interactions.
If only the triple sum transition is resonant, we can approximate all other resonance ($\Delta$) terms as constant and arrive at an expression similar to \autoref{E:chi1}:\cite{Morrow_Wright_2018}
\begin{equation}
	\chi^{(3)}(\omega_{123}) \propto \sum_{a, g} \frac{\mu_{ag}^4}{\Delta_{ag}^{123}} \label{E:chimuJ}.
\end{equation}

We now consider how the linear and third-order susceptibilities dictate the reflectance and TSF response, respectively.
Both relations are formulated using Maxwell's equations via continuity relations (boundary conditions) between the incident, reflected, and transmitted fields.
For ease of comparison, we will analyze the simple limit of an extremely thin film (thickness much less than the wavelength of light) on a transparent substrate.
We also restrict consideration to normal incidence.
Including thickness and angular dependence is straightforward but needlessly complex for our intent of illustrating qualitative differences between methodologies.
These conditions are reasonable for many of the samples and experiments we consider here.

With these conditions, the reflectance, $R$, is given by\cite{Falkovsky_2008, Sie_Gedik_2015}
\begin{equation}
	R \equiv \frac{I_\text{reflected}}{I_1} = \frac{\left(1 - A\right)^2 + B^2}{\left(1 + A\right)^2 + B^2}, \label{E:R}
\end{equation}
where
\begin{align}
	A &\equiv n_s + \frac{\omega_1 \ell}{c} \text{Im}\left[\chi^{(1)}\right], \\
	B &\equiv \frac{\omega_1 \ell}{c} \text{Re}\left[\chi^{(1)}\right],
\end{align}
in which $\ell$ is the film thickness (propagation length), $n_s$ is the substrate refractive index, $c$ is the speed of light in vacuum, and $I_j$ is the intensity of the $j$\textsuperscript{th} electric field.
Note that when $\frac{\omega_1 \ell}{c} \left|\chi^{(1)}\right| \ll n_s$, \autoref{E:R} is primarily determined by the substrate refractive index (large background reflectance).
For example, taking a nominal $\chi^{(1)}$ value of $\sim 1$, and a few-layer thickness, $\ell \sim 10$ nm, we calculate $\frac{\omega_1 \ell}{c} \left|\chi^{(1)}\right| \approx 0.1$ while $n_s \approx 1.45$ (both for excitation colors near the band edge of TMDCs), so the thin film limit will be appropriate for several samples considered in this work.

Expanding \autoref{E:R} and keeping only terms linear in $\chi^{(1)}$, shows that the imaginary component of the thin film susceptibility is responsible for contrast from the substrate background:
\begin{align}
R 	&\approx \frac{
		\left(1 - n_s\right)^2
		+2(n_s - 1)\frac{\omega_1 \ell}{c}\text{Im}\left[\chi^{(1)}\right]
	}{
		\left(1 + n_s\right)^2
		+2(1 + n_s)\frac{\omega_1 \ell}{c}\text{Im}\left[\chi^{(1)}\right]
	} \label{E:R_thin_limit_1}.
\end{align}
\autoref{E:R_thin_limit_1} can be further simplified by Taylor expansion around $\frac{2\omega_1 \ell}{c}\text{Im}\left[\chi^{(1)}\right]=0$:
\begin{align}
R	&\approx R_0 -\left(\frac{R_0}{1+n_s} + \frac{1-n_s}{\left(1+n_s\right)^2}\right)\frac{2\omega_1 \ell}{c}\text{Im}\left[\chi^{(1)}\right], \label{E:R_thin_limit}
\end{align}
where $R_0 \equiv \frac{\left( 1 - n_s \right)^2}{\left( 1 + n_s \right)^2}$ is the reflectance of the substrate-air interface.

TSF emission, or non-linear frequency conversion in general, is qualitatively different from reflectance (or transmittance) because the TSF wave originates from inside the thin film.\footnote{TSF emission from the substrate is also possible, but in practice this contribution is negligible compared to TMDC thin films when measuring in the reflective direction c.f. the SI of \textcite{Morrow_Wright_2018}}
This difference brings two important consequences to the measured beam: (1) TSF emission is dark in regions where the thin film is not present, and (2) the continuity relations are acutely sensitive to the thin film non-linear polarization, rather than an incident field.\cite{Bloembergen_Pershan_1962}
For the aforementioned thin film conditions, the TSF output intensity satisfies the proportionality
\begin{align}
	\frac{I_{\textrm{TSF}}}{I_1 I_2 I_3} \propto \left|\chi^{(3)}\right|^2 (\omega\ell)^2. \label{E:ITSF}
\end{align}
Unlike reflectance, thin film TSF emission obeys the same $\chi^{(3)}$ scaling as the thick film emission case,\cite{Boyle_Wright_2013} where the film thickness is larger than or close to the wavelength of light, but phase mismatch effects are still small.

\subsection{Pump-TSF-probe and TR spectroscopy}

We now consider how the different nature of the reflectance and TSF probe result in different, yet similar, pump-probe measurements.
For both linear and non-linear probes, we can describe the pump-induced susceptibility as a perturbation to the unpumped susceptibility:
\begin{equation}
	\chi^{(n)}_\text{pumped} = \chi^{(n)}_\text{unpumped} + \textrm{d}\chi^{(n)}, \label{E:dchi}
\end{equation}
where $\textrm{d}\chi^{(n)} = \chi^{(n+2)} I_\text{pump}$ is the small pump-induced perturbation.
Pump-probe methodologies often look at relative changes in the probe:
\begin{align}
\textrm{signal metric} = \frac{X_\text{pumped} - X_\text{unpumped}}{X_\text{unpumped}} \label{E:dXX}
\end{align} 
in which $X$ is the probed quantity.
This normalization generally allows for electric fields which are not spectrally flat to be used as a probe.

Using reflectance as our probe (\autoref{E:R_thin_limit}) gives a transient response of
\begin{equation}
	\frac{\Delta R}{R} \approx \frac{-1}{R}
	 \left(\frac{R_0}{1+n_s} + \frac{1-n_s}{\left(1+n_s\right)^2}\right)\frac{2\omega_1 \ell}{c}\text{Im}\left[\textrm{d}\chi^{(1)}\right].
\end{equation}
This expression shows that our signal metric scales as $\text{Im}\left[\textrm{d}\chi^{(1)}\right]$ which is the same as transient transmittance in a bulk sample (see Appendix \ref{A:TT} for a derivation).
In other words, in the extremely thin film limit, transient reflectance will have lineshapes which are intuitive to those who are used to interpreting bulk transient transmittance (absorption) measurements.
The intuitive correspondence between transient reflectance and transient transmittance spectrosocpies will break down as $\frac{\omega_1 \ell}{c} \left|\chi^{(1)}\right|$ increases---thick samples require a full Fresnel analysis to understand the transient reflectance lineshapes. 

With TSF intensity as our probe, we use \autoref{E:dchi} and \autoref{E:ITSF} to arrive at
\begin{equation}
	\frac{\Delta I_{\textrm{TSF}}}{I_{\textrm{TSF}}}
		= \frac{
			\left|\textrm{d}\chi^{(3)}\right|^2
			+ 2 \left|\textrm{d}\chi^{(3)}\right| \left|\chi^{(3)}\right| \cos (\textrm{d}\theta)
		}{
			\left|\chi^{(3)}\right|^2
		},\label{E:dII_chi}
\end{equation}
where we have used phasor representations of the susceptibilities: $\chi^{(3)} \equiv \left|\chi^{(3)}\right| e^{i\theta}$ and $\textrm{d}\chi^{{(3)}} \equiv \left|\textrm{d}\chi^{(3)}\right| e^{i(\theta + \textrm{d}\theta)}$, in which $\theta$ can be dependent on probe frequency.
If $\left|\textrm{d}\chi^{(3)}\right| \ll \left|\chi^{(3)} \cos (\textrm{d}\theta) \right|$ we can write
\begin{equation}
	\frac{\Delta I_{\textrm{TSF}}}{I_{\textrm{TSF}}} \approx 2 \left|\frac{\textrm{d}\chi^{(3)}}{\chi^{(3)}}\right| \cos (\textrm{d}\theta), \label{E:dIsim}
\end{equation}
If the pump changes only the amplitude of $\chi^{(3)}$ ($\textrm{d}\theta = 0, \pi)$, the relative change in TSF intensity tracks the relative change in susceptibility.
However, if the pump also changes the phase, the amplitude changes can be suppressed.
Note that in the case of a $\pi/2$ phase shift, our assumption behind \autoref{E:dIsim} is invalid.
It is important, then, to understand when $\textrm{d}\theta$ can be large.
In general, $\theta$ changes rapidly near resonances; if pump induced changes shift or broaden a resonance to an extent similar to its linewidth, $\textrm{d}\theta$ will strongly influence the pump-TSF-probe spectrum.
In the absence of dramatic resonance changes, lineshapes will closely approximate $\textrm{d}\chi^{(3)} / \chi^{(3)}$.

\begin{figure}[!htbp]
	\centering
	\includegraphics[width=\linewidth]{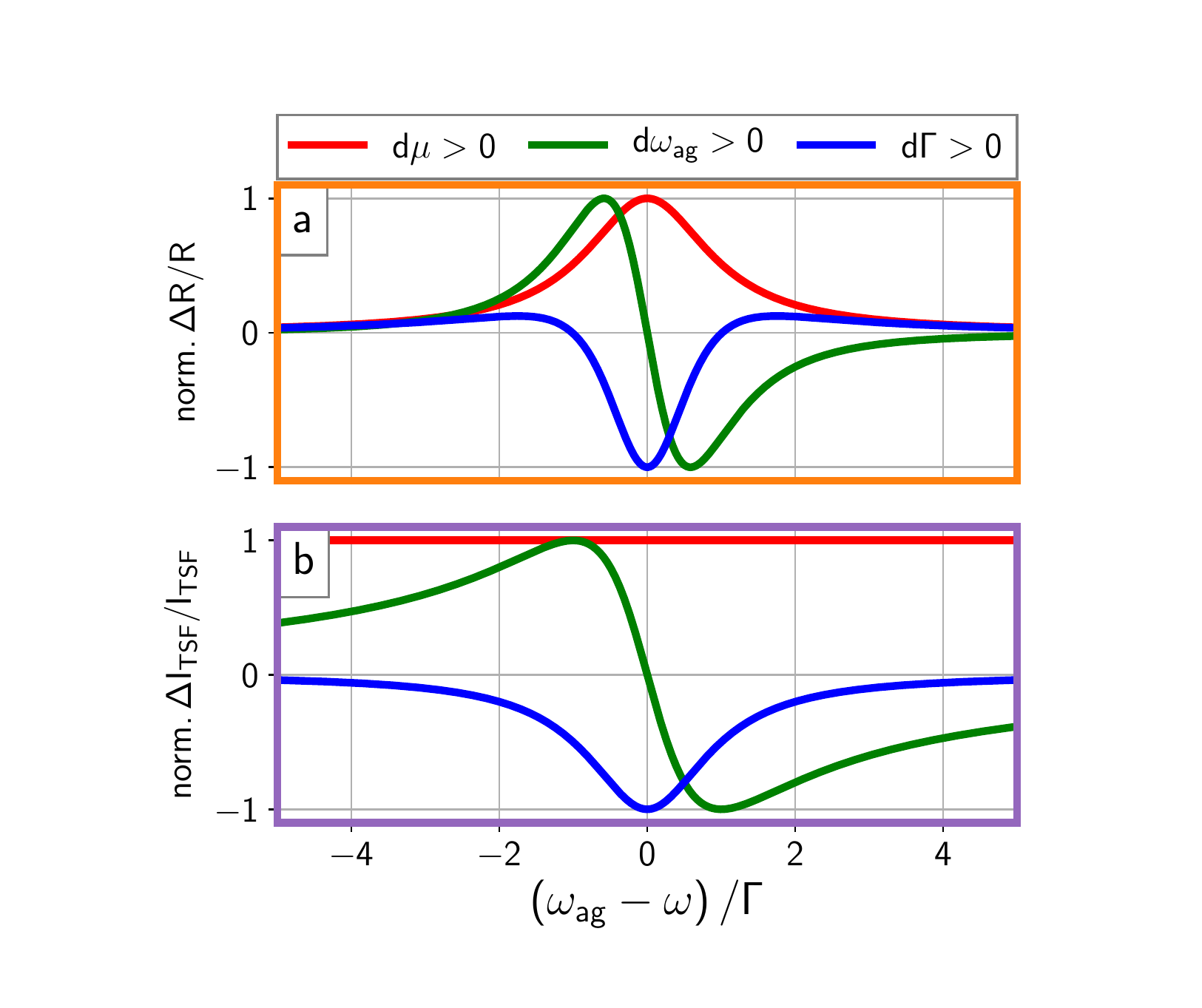}
	\caption{
		Calculated transient lineshapes for a single, complex Lorentzian resonance (c.f. \autoref{E:chi1} and \autoref{E:chimuJ}). (a) the transient-reflectance spectrum. (b) the transient-TSF spectrum.
		The spectra are produced by perturbing $\mu$, $\omega_{ag}$, or $\Gamma$ by a factor of $10^{-5}$.\label{fig:lineshapes}}
\end{figure}

To anticipate the spectra of each technique, it is useful to consider the case of a single Lorentzian resonance perturbed by the pump.
For small perturbations we can construct $\textrm{d}\chi^{(n)}$ using the total derivative
\begin{align}
\textrm{d}\chi^{(n)} &= \frac{\partial\chi^{(n)}}{\partial \mu}\textrm{d}\mu
+ \frac{\partial\chi^{(n)}}{\partial \omega_{ag}}\textrm{d}\omega_{ag}
+ \frac{\partial\chi^{(n)}}{\partial \Gamma}\textrm{d}\Gamma \label{E:total_deriv}. 
\end{align}
In the appendices we derive analytical expressions for the lineshapes expected from transient-TSF and transient-transmittance spectroscopies in this single resonance limit. 
Numerically calculated spectra are shown in \autoref{fig:lineshapes} for three different types of perturbations: 
\begin{itemize}
	\item An increase in the transition dipole, $\textrm{d}\mu > 0$. For an excitonic transition, state-filling and Coulomb-screening will usually lead to a decrease in the transition dipole. Note that changes in state density will cause the same lineshape as transition dipole changes. 
	\item An increase in the resonance frequency, $\textrm{d}\omega_{ag} > 0$. For an excitonic transition, bandgap renormalization or Coulomb-screening can lead to decreases or increases in the resonance freuquency.
	\item An increase in the dephasing rate, $\textrm{d}\Gamma >0$. An increase in particle-particle scattering rates due to pump-excited carriers can cause the dephasing rate of a transition to increase.
\end{itemize}

The transient-reflectance spectra (see \autoref{fig:lineshapes}a) are easily interpreted because changes in $\text{Im}\left[\chi^{(1)}\right]$ correlate with changes in absorptive cross-section (\autoref{E:R_thin_limit}).
Interpretation of $\Delta R / R$ line shapes mirrors the traditional interpretation of differential transmission, $\Delta T / T$, for bulk samples obeying Beer's law.
Increasing the dipole strength (red line) increases reflectance (positive $\Delta R /  R$), with a line shape mirroring the unpumped transition.
Resonance red-shifts (green line) increase reflectance to the red and decreases reflectance to the blue. 
Line shape broadening (blue line) decreases reflectance in the center of the resonance but increases reflectance on the wings.

The transient-TSF lineshapes (\autoref{fig:lineshapes}b) have similar interpretations.
Increases in transition dipole (red line) increases TSF (positive $\Delta I / I$).
Unlike reflectance, the increase results in a constant $\Delta I / I$ offset.
This is because the unpumped $I_\text{TSF}$ lineshape has no background from substrate and so its lineshape is sharply peaked and matches that of $\Delta I$. 
Line shape broadening (blue line) and blue-shifting (green line) again mirror the behavior of $-\Delta T / T$, but the $\Delta I / I$ line shapes are noticeably broader than $\Delta R / R$.
Since TSF is sensitive not only to imaginary component, but also the dispersive real component of $\chi^{(3)}$ (\autoref{E:dIsim}), the resulting lineshapes are intrinsically broader.
In general, for the same dephasing rate, the transient-TSF lineshapes are broader than the transient-transmittance and transient-reflectance lineshapes.

\section{Experimental}\label{S:Experimental}

\subsection{Ultrafast measurements}

Our experimental setup uses an ultrafast oscillator seeding a regenerative amplifier (Spectra-Physics Tsunami and Spitfire Pro, respectively) to produce $\sim$35 fs pulses centered at 1.55 eV at a 1 kHz repetition rate. 
The amplifier output separates into three arms. 
Not all arms are used in every experiment.
Two arms are optical parametric amplifiers (Light-Conversion TOPAS-C) which create tunable pulses of light from $\sim$0.5 to $\sim$2.1 eV with spectral width on the amplitude level of FWHM $\approx$ 46 meV, absorptive filters and wire grid polarizers are used to isolate light of the desired color.\footnote{
	A crucial filter for our TSF probe experiments is a 1000 nm longpass filter (ThorLabs FGL1000M) which ensures no visible light from secondary OPA processes reach the sample. Notably, double polished silicon was not a suitable filter because it created non-negligible THG as well as lossy transmission.} 
The third arm frequency doubles the output of the amplifier to create pulses centered at 3.1 eV in a $\beta$-barium-borate crystal. 
Each arm has its own mechanical delay stage and optical chopper. 
All pulses are then focused onto the sample with a 1 m focal length spherical mirror. 
The spatially coherent output (either the reflected probe or the triple sum of the probe) is isolated with an aperture in the reflected direction (sometimes referred to as an \emph{epi} experiment), focused into a monochromator (Horiba Micro-HR) and detected with a thermoelectrically cooled photomultiplier tube (RCA C31034A). 
A dual-chopping routine is used to isolate the desired differential signal.\cite{Furuta_Wada_2012}
The color-dependent time-of-flight for each arm is corrected by offsetting the mechanical delay stages for each combination of pump and probes colors. 
We use a reflective geometry for our TSF measurements in order to minimize phase-mismatch effects.\cite{Morrow_Wright_2017, Handali_Wright_2018}
Unless otherwise noted, the pump fluence in these measurements is $\sim$100 $\mu$J/cm\textsuperscript{2}. 
The visible probe beam for the reflectance-probe experiments has a fluence of $\sim$2 $\mu$J/cm\textsuperscript{2} while the NIR probe lasers for the TSF-probe experiments have a fluence of $\sim$1000 $\mu$J/cm\textsuperscript{2}.
All beams are hundreds of microns wide at the sample.
All raw data, workup scripts, and simulation scripts used in the creation of this work are permissively licensed and publicly available for reuse.\cite{OSF_Morrow_2019_A} 
Our acquisition\cite{PyCMDS} and workup\cite{Thompson_Wright_2019} software are built on top of the open source, publicly available Scientific Python ecosystem.\cite{Jones_2001, vanderWalt_Varoquaux_2011, Hunter_2007} 

\subsection{Sample preparation and characterization}

Polycrystalline MoS\textsubscript{2} (WS\textsubscript{2}) films were prepared by first e-beam evaporating 2 nm of Mo (W) onto a fused silica substrate and subsequent sulfidation in a tube furnace at 750 $^{\circ}$C for 10 (30) minutes.\cite{Czech_Wright_2015}
Note that this exact MoS\textsubscript{2} thin film sample was previously explored in \textcite{Morrow_Wright_2018}.
Following the methods detailed in \textcite{Zhao_Jin_2019_arxiv}, WS\textsubscript{2} (MoS\textsubscript{2}) nanostructure samples on 300 nm SiO\textsubscript{2}/Si substrates were prepared using water vapor assisted chemical vapor transport growth by heating 100 mg WS\textsubscript{2} (MoS\textsubscript{2}) powder to 1200 $^{\circ}$C at 800 torr in a tube furnace in which water vapor was produced by heating  1 g CaSO\textsubscript{4}$\cdot$2H\textsubscript{2}O powder to 150 $^{\circ}$C (120 $^{\circ}$C) using heating tape wrapped around the tube furnace. 100 sccm argon was used as the carrier gas during the reaction.

\begin{figure}[!htbp]
	\centering
	\includegraphics[width=\linewidth]{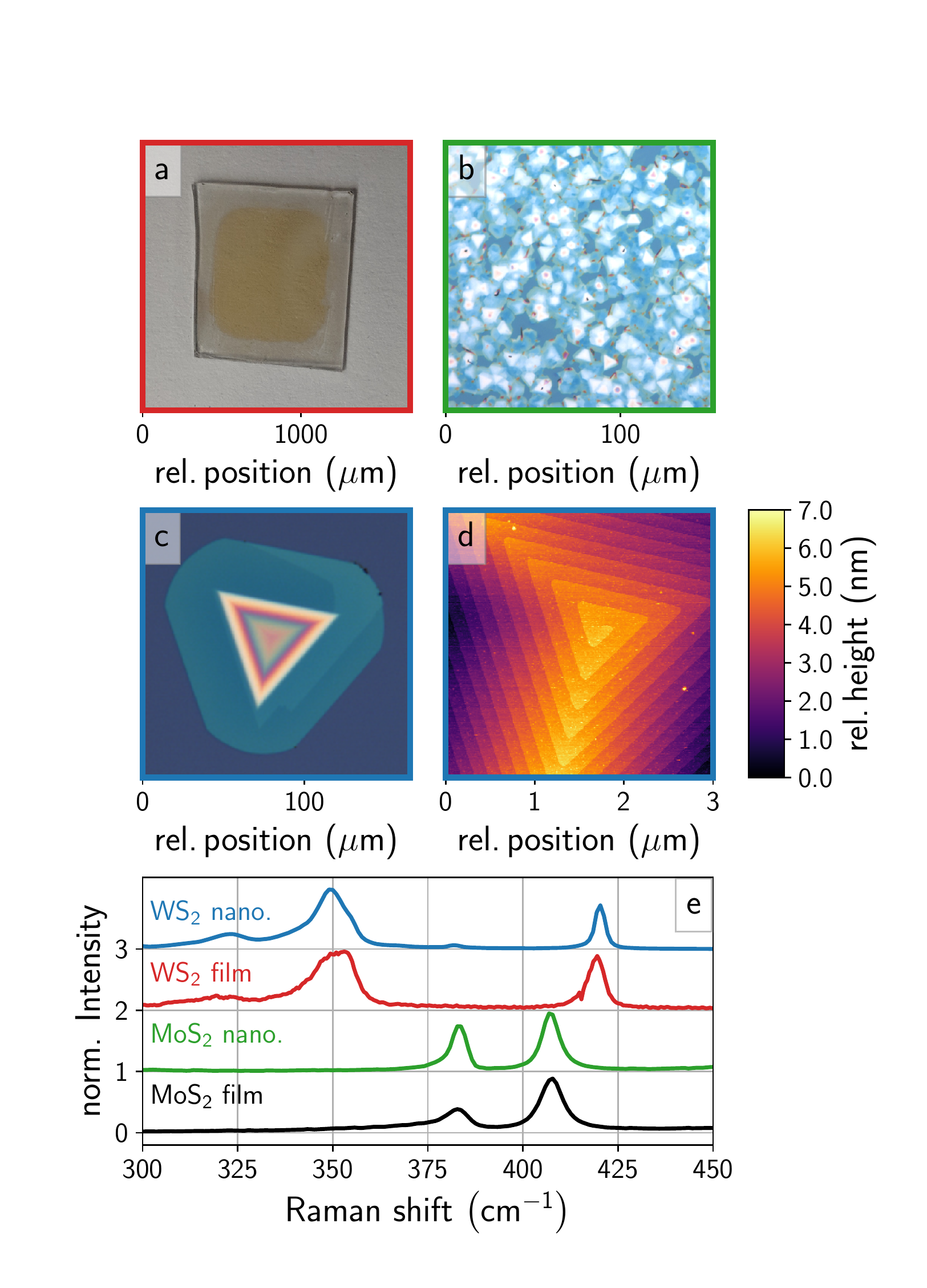}
	\caption{
		TMDC Sample characterization. 
		(a) a photograph of the WS\textsubscript{2} film explored in this work.
		(b) an optical microscope image of the MoS\textsubscript{2} nanostructures explored in this work.
		(c) optical microscope and (d) atomic force microscope image of the WS\textsubscript{2} nanostructure explored in this work.
		(e) Raman spectra of each sample; the maximum of each Raman spectrum is normalized to 1 and then offset for ease of comparison.}
	\label{fig:characterization}
\end{figure}

\autoref{fig:characterization} shows optical microscope (Olympus BX51M) images, atomic force microscope (Agilent 5500) data, and Raman spectra (LabRAM Aramis, Confocal Raman/PL Microscope, 2.33 eV excitation) of the samples.
The Raman spectra show the common $\textrm{E}^1_{2\textrm{g}}$ and $\textrm{A}_{1\textrm{g}}$ modes seen in WS\textsubscript{2} and MoS\textsubscript{2} at the frequencies expected for many-layer to bulk morphologies.\cite{Lee_Ryu_2010, Li_Baillargeat_2012, Berkdemir_Terrones_2013}
The polycrystalline thin films ($\sim$10 nm thick) are continuous, flat, and smooth samples that are much larger than the spot size of our lasers (see \autoref{fig:characterization}a). 
Each MoS\textsubscript{2} nanostructure (\autoref{fig:characterization}b) is a few microns wide and sparsely scattered across the substrate.
The nanostructures exhibit a wide range of morphologies from screw-dislocation spirals to stacked plates.
The WS\textsubscript{2} nanostructure (\autoref{fig:characterization}c and d) is a single screw-dislocation spiral which is 84 nm ($\sim$120 layers) thick and $\sim$150 $\mu$m wide.
TMDC screw-dislocation spirals are known to have excellent optical harmonic generation abilities.\cite{Shearer_Jin_2017, Fan_Pan_2018, Fan_Pan_2017, Zhang_Yang_2014} 
Note that the perceived colors of the nanostructures in \autoref{fig:characterization}b and \autoref{fig:characterization}c are due to thin-film interference effects from the combination of the pyramid nanostructures, which have a large change in height across the structure, and the SiO\textsubscript{2}/Si substrates; this class of effects have previously been explored for monolayers and nanostructures.\cite{Zhang_Dai_2015, Benameur_Kis_2011, Blake_Geim_2007}

\section{Results and Discussion}\label{S:Results}

\subsection{MoS\textsubscript{2} thin film: transient-TSF}

\begin{figure}[!htbp]
	\centering
	\includegraphics[width=\linewidth]{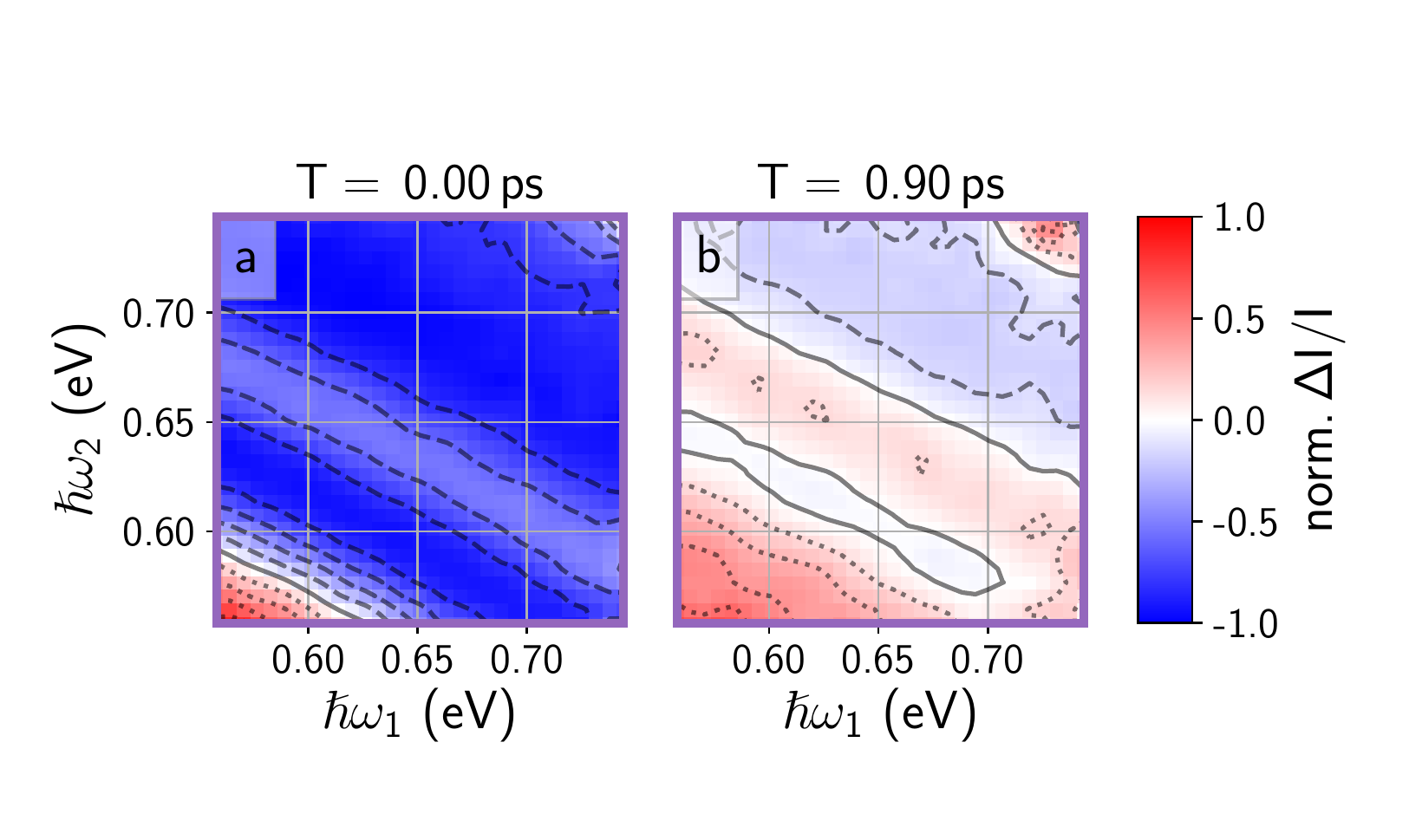}
	\caption{
		Pump-TSF-probe spectra of an MoS\textsubscript{2} thin film at 0 ps (a) and 0.90 ps (b) delay between pump and probe interactions. 
		In both frames $\hbar\omega_{\text{pump}}=3.1\text{ eV}$ with a fluence of 120 $\mu$J/cm\textsuperscript{2}, $\omega_m = \omega_1 + 2\omega_2$, and $\vec{k}_{\text{out}} = -\left(\vec{k}_1 + 2\vec{k}_2 \right)$. }
	\label{fig:ptsfp}
\end{figure}

We first examine the effect of a pump on the multidimensional TSF spectrum, in which $\omega_1$ and $\omega_2$ are independently scanned.
The output frequency of the TSF probe, $\omega_m = \omega_1 + 2\omega_2$, covers the range of the A and B excitons (1.65 - 2.25 eV).
We explore this dependence with a MoS\textsubscript{2} thin film.
In this film, the unpumped multidimensional spectra exhibit singly resonant features that depend only on the output triple sum frequency (e.g. \autoref{E:chimuJ}).\cite{Morrow_Wright_2018}
There are no cross peaks in the unpumped TSF spectrum because MoS\textsubscript{2} A and B excitons do not have the correct symmetry for our excitation beams to couple together.
Like the unpumped spectrum, we found that the pump-TSF-probe spectra depends only on the triple sum frequency, regardless of pump-probe delay time.
Pump-TSF-probe spectra of the MoS2 thin film at two different delays are shown in \autoref{fig:ptsfp} ($\hbar\omega_{\text{pump}}=3.1\text{ eV}$).
At both delay times, all features run along lines of constant output color (slope of -1/2).
We explored the multidimensional probe spectra at other frequency and $T$ combinations (output energies up to 3 eV and population times up to 100 ps); all features found are defined solely by the output color.

Given the similarities in band structure, we expect this result to be general to all TMDCs.
The simplicity of the TSF and pump-TSF-probe spectra motivate the use of \autoref{E:chimuJ} and its associated pump-THG-probe analysis which was discussed in the Theory section.
Since the output color seems to uniquely determine the observed spectra, we restrict ourselves to the case $\omega_1 = \omega_2 = \omega_m/3$ (pump-THG-probe) for the rest of this work.
We will understand the lineshapes present in \autoref{fig:ptsfp} by understanding the lineshapes of the pump-THG-probe spectroscopy presented in the next section.

\subsection{MoS\textsubscript{2} thin film: transient-THG and transient-reflectance}

\begin{figure}[!htbp]
	\centering
	\includegraphics[width=\linewidth]{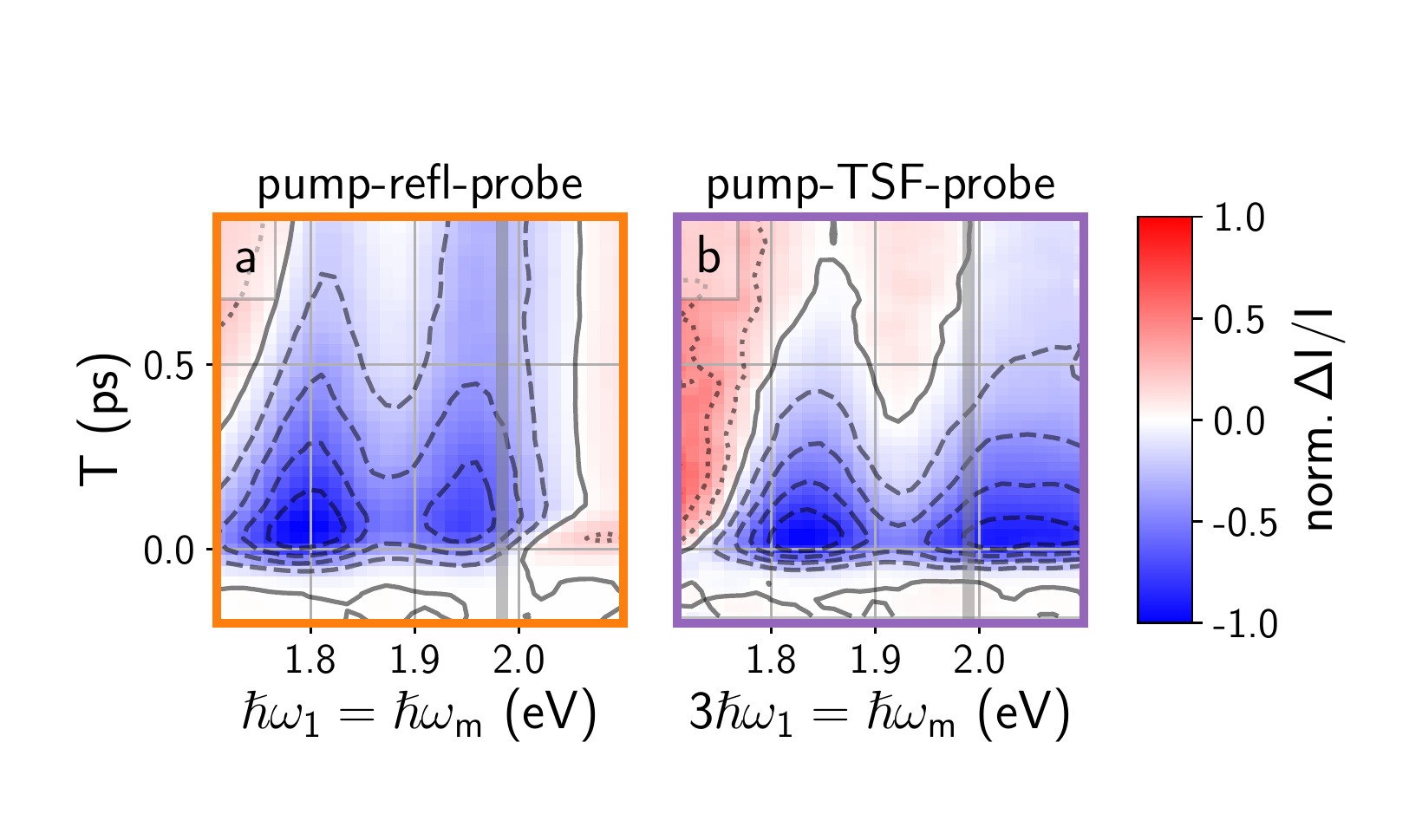}
	\caption{
		Comparison of transient-reflectance spectroscopy (a) to transient-TSF spectroscopy (b) for a MoS\textsubscript{2} thin film.
		In both frames $\hbar\omega_{\text{pump}}=1.98\text{ eV}$, as indicated by the vertical gray line.
		Each plot has its own colormap extent, red (dotted contours) signifies $\Delta I/ I > 0$, white (solid contour) signifies $\Delta I/ I = 0$, and blue (dashed contours) signifies $\Delta I/ I < 0$.}
	\label{fig:wigner}
\end{figure}

\autoref{fig:wigner} shows both the pump-reflectance-probe (left) and the pump-TSF-probe (right) response of the MoS\textsubscript{2} thin film with pump excitation at the B exciton. 
Note that horizontal $3\omega_1$ slices through \autoref{fig:wigner}b are almost equivalent to the diagonal, $\omega_1 = \omega_2$ slices of \autoref{fig:ptsfp}; they differ only in the use of different pump colors.
The TR and pump-THG-probe spectra are qualitatively similar
Our analysis in the Theory section indicates that phenomena like shifting and broadening will lead to similar lineshapes between the two spectroscopies while state density and dipole decreases will look different between the two spectroscopies---so our observed response is likely due to shifting and broadening of the underlying excitonic resonances.
In both spectra, the measured intensity at the A and B excitons decreases when the pump is on ($\Delta I / I < 0$).
The intensity decreases dominate at $T=0$, then decay over $\sim$500 fs to form spectra that undulate between positive and negative values. 
These undulating spectra persist for several picoseconds (data not shown).  

The minima of the transient-THG spectrum are blue-shifted relative to the corresponding minima observed in the transient-reflectance spectrum, but roughly agree with the peak positions of the unpumped THG spectrum (\autoref{fig:1D}).
The A and B peaks of the unpumped THG spectrum are blue-shifted by $\sim$50 meV compared to the absorption/reflection spectrum.
We cannot explain why the maxima of the THG and absorption/reflection spectra are different by $\sim$50 meV, but we note that \textcite{Wang_Zhao_2013} also observed this same unexplained blue-shift in their THG measurements. 

\begin{figure}[!htbp]
	\centering
	\includegraphics[width=\linewidth]{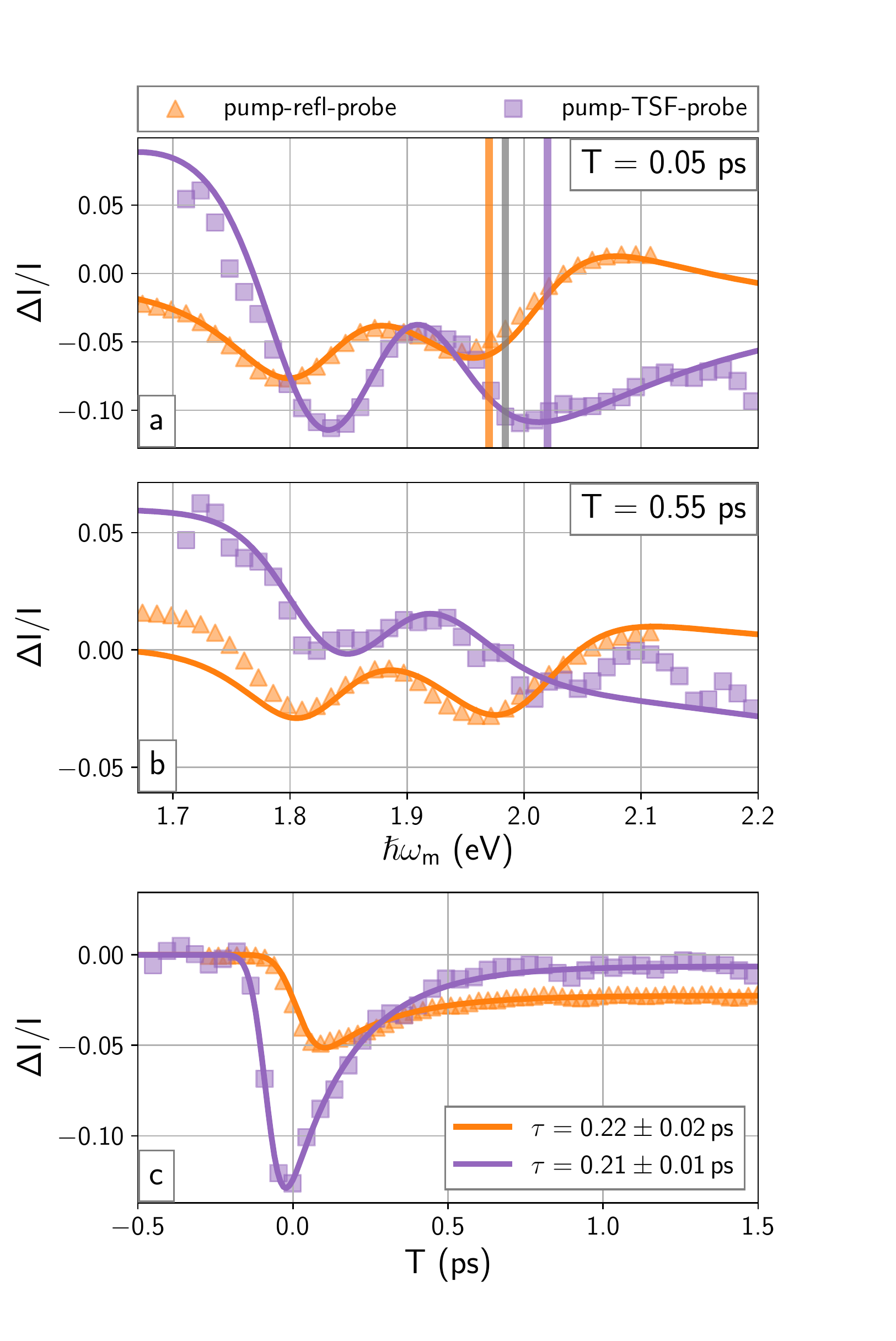}
	\caption{
		Comparison of spectral and temporal lineshapes with $\hbar\omega_{\text{pump}}=1.98\text{ eV}$ (gray vertical line).
		Spectral lineshapes in (a) and (b) are acquired with delay times of 0.05 and 0.55 ps, respectively.
		Dynamics in (c) are acquired at probe energies indicated by the vertical lines in (a) (1.97 and 2.02 eV for pump-refl-probe and pump-TSF-probe, respectively).
		Solid lines in each plot are calculated from the models described in the main text and Appendix \ref{A:lineshapes}.}
	\label{fig:fitting}
\end{figure}

To understand the spectral and temporal information in \autoref{fig:wigner}, we examine transients at fixed probe frequencies and spectra at fixed time delays in \autoref{fig:fitting}.
For both cases, we use simple models to ensure standard physical arguments can explain our observations.
The specifics of the spectral lineshape model (results shown as solid lines in \autoref{fig:fitting}a and \autoref{fig:fitting}b) are discussed in Appendix \ref{A:lineshapes}.

We first consider the spectral slices.
In both spectroscopies, the lineshape at $T \approx 0$ (\autoref{fig:fitting}a) can be explained by a $\sim$10 meV redshift, slight broadening, and slight amplitude decreases of the A and B resonances.
Appendix \ref{A:lineshapes} details the parameters used to generate the solid lines in \autoref{fig:fitting}a and \autoref{fig:fitting}b.
A short time after excitation, $T = 0.55 \text{ ps}$, the spectra are defined by a few meV redshift, no broadening, and no amplitude decrease.
In order to describe the pump-TSF-probe lineshape in \autoref{fig:fitting}b we incorporated an additional ESA background.
We attribute the additional ESA to excitation of near band edge carriers (conduction band electrons, valence band holes, or excitons) upon pump photoexcitation and subsequent relaxation.
We attribute the redshift to carrier-induced bandgap renormalization (BGR), which was previously predicted and observed by many in monolayer TMDCs).\cite{Pogna_Prezzi_2016, Liu_Zhu_2019, Chernikov_Heinz_2015, Steinhoff_Gies_2014, Meckbach_Koch_2018}
The lineshape broadening upon excitation is common in semiconductors and has been previously observed by refs. \cite{Sim_Choi_2013, Cunningham_Jonker_2017} in TMDCs.
The amplitude decrease is likely due to state/band filling, in this simple model, we cannot distinguish between changes in transition density and transition dipole.

Dynamics were described by an exponential decay and a static offset:
\begin{equation}
	\frac{\Delta I}{I}(T) = \left(A_0  \exp{\left(-\frac{T}{\tau}\right)} + A_1\right) \Theta\left(T-t_0\right),  \label{E:fit}
\end{equation}
in which $\Theta$ is the Heaviside step function.
We convolve \autoref{E:fit} with an instrument response function, which we approximate as Gaussian.
Results are shown as solid lines in \autoref{fig:fitting}c).
Both spectroscopies exhibit time constants of $\sim$200 fs (\autoref{T:dynamics}).\footnote{Because these lineshapes are not merely caused by amplitude  changes ($J$ or $\mu$), fitting different probe colors results in different decay rates, with $\tau$ up to 0.4 ps.}
Dynamics on this timescale have previously been attributed to several mechanisms, including carrier trapping,\cite{Cunningham_Hayden_2016, Shi_Huang_2013, Schiettecatte_Hens_2019} carrier-carrier scattering,\cite{Sim_Choi_2013, Tsokkou_Banerji_2016} carrier-phonon scattering,\cite{Nie_Loh_2014, Kumar_Zhao_2013_B, Nie_Loh_2015} free-carrier screening effects,\cite{Cunningham_Jonker_2017, Ceballos_Zhao_2016} and exciton formation.\cite{Ceballos_Zhao_2016}
The longer dynamic in \autoref{fig:fitting}c, which we treat as an offset, $A_1$, has been observed by others.\cite{Shi_Huang_2013, Ceballos_Zhao_2016} 

\begin{table}[!htbp]
	\caption{\label{T:fitresults} Results from fitting \autoref{E:fit} to the dynamics shown in \autoref{fig:fitting}b. FWHM = full width at half maximum of the instrument response function.}
	\begin{ruledtabular}
		\begin{tabular}{lll}\label{T:dynamics}
			& pump-refl-probe & pump-TSF-probe \\
			\hline
			$\hbar\omega_m$ (eV)  				& 1.97            & 2.02           \\
			$\tau$ (ps)  & 0.22 $\pm$ 0.02   	 & 0.21 $\pm$ 0.01   \\
			FWHM (ps) & 0.125 $\pm$ 0.009 		 & 0.095 $\pm$ 0.006
		\end{tabular}
	\end{ruledtabular}
\end{table}

\begin{figure}[!htbp]
	\centering
	\includegraphics[width=\linewidth]{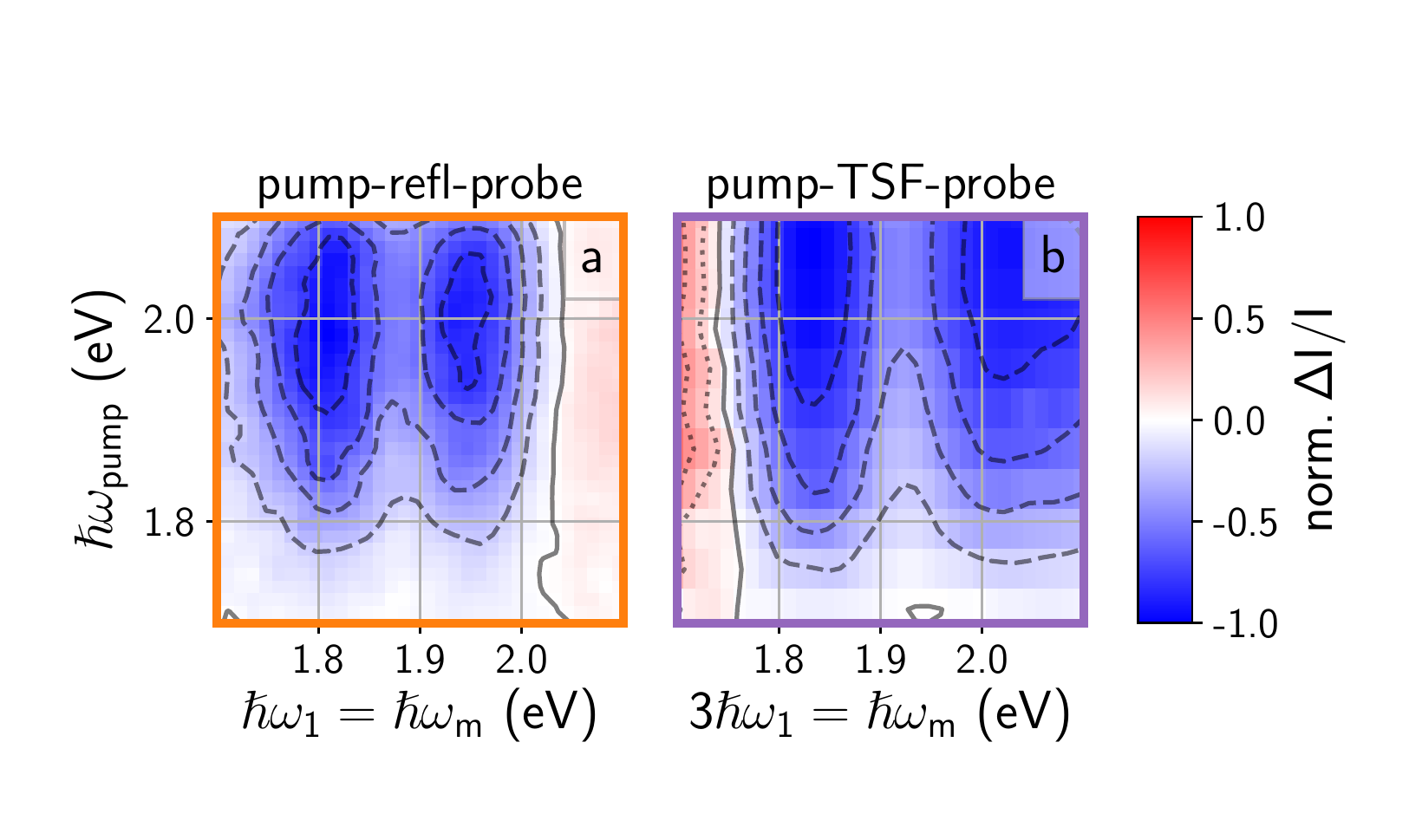}
	\caption{
		Comparison between transient-reflectance spectroscopy (a) and transient-TSF spectroscopy (b) of a MoS\textsubscript{2} thin film. 
		In both frames $T=50\text{ fs}$. }
	\label{fig:freq}
\end{figure}

\autoref{fig:freq} shows the response from both TR and transient-THG spectroscopies for different combinations of pump and probe frequencies when $T=50 \text{ fs}$.
\autoref{fig:freq}a is a very similar measurement to refs. \cite{Czech_Wright_2015, Singh_Li_2014, Moody_Li_2015, Singh_Li_2016, Hao_Li_2016, Hao_Moody_2016, Hao_Moody_2017, Guo_Fleming_2018} where ``traditional'' coherent multidimensional spectroscopies were accomplished on TMDCs using a single electric field interaction as a probe; this measurement similarity is not the case for \autoref{fig:freq}b in which TSF acts as the probe with three electric field interactions.
Nevertheless, both of our spectroscopies in \autoref{fig:freq} have a similar dependence on the pump frequency---this can be seen by comparing vertical slices of \autoref{fig:freq}a and b (these horizontal slices are analogous to horizontal slices of \autoref{fig:wigner}.).\footnote{the decrease in $\Delta I/I$ at high pump frequencies in the TR experiment \autoref{fig:freq}a is likely caused by a decrease in the $I_\text{pump}$ at those frequencies.  
The two spectra in Figure \ref{fig:freq} were collected at different times and do not share the same pump intensity spectrum.}
The lineshapes of our two spectroscopies change smoothly as a function of $\hbar\omega_{\text{pump}}$---there are no distinct contributions from the A or B resonances along the pump axis.
The lack of structure along the pump axis mirrors the results of transient grating measurements on a MoS\textsubscript{2} thin film.\cite{Czech_Wright_2015}
The general insensitivity to pump color suggests band gap renormalization (BGR) is a large contributor to the pump-induced changes.
BGR is determined by Coulomb interactions, which are less sensitive to the explicit electron and hole occupation than, for instance, Pauli blocking effects.

Conversely, small, but noticeable, dependencies on $\omega_{\text{pump}}$ suggest secondary contributions to the TR and transient-TSF spectra.
For instance, when $\hbar\omega_{\text{pump}} \approx 1.8 \text{ eV} \approx \hbar\omega_{\text{A}}$, the decreases in intensity at $\omega_\text{out} = \omega_A$, are $\sim 15 \%$ larger than at $\omega_\text{out} = \omega_B$ for both TR and pump-TSF-probe. 
When $\hbar\omega_{\text{pump}} \approx \hbar\omega_{\text{B}}$, however, both A and B have similar decreases in intensity.
We believe band/state filling (bleaching) can account for the observed asymmetries in the decreases in intensity.
The MoS\textsubscript{2} valence band is energetically split for different hole spins, but the conduction band is energetically degenerate for electron spins (cf. inset in \autoref{fig:1D}).
The A transition and B transition share neither valence nor conduction bands, so state/band filling is not shared between transitions.
When the pump is resonant with the A transition, the bleach of the B transition is not direct and is expected to be smaller, in agreement with our measurements.
Some bleaching is allowed through intervalley scattering, but valley depolarization measurements suggest these timescales are much longer than our pump probe delay time (50 fs) and is thus not significant.\cite{Mahmood_Gedik_2017,Yang_Crooker_2015, Moody_Xu_2016} 
When the pump is resonant with the B transition, it will also directly excite hot excitons or free electons/holes from the A band, which explains why the effects on the A and B transitions are similar for these pump colors.

\textcite{Guo_Fleming_2018} also found asymmetries in the cross peaks of their multidimensional spectra of monolayer MoS\textsubscript{2} at 40 K. 
They attribute the asymmetric cross-peaks and their dynamics to be due to an exchange interaction between A and B excitons. 
This effect does not explain our results because the exchange interaction requires simultaneous populations of A and B excitons, which is not the case for $\omega_\text{pump} \approx \omega_A$. 
	
\subsection{MoS\textsubscript{2} and WS\textsubscript{2} nanostructures}

In this section we investigate the effects of sample morphology on pump-TSF-probe spectroscopy by comparing all the samples shown in \autoref{fig:characterization}.
We first compare spectra of the previously discussed thin film with an ensemble of spiral nanostructures grown via a screw dislocation driven growth method (\autoref{fig:characterization}b).
The goal of this comparison is to broadly demonstrate that the spectra and dynamics observed with transient-TSF are sensitive to the specifics of sample morphology.
We then compare the transient-TSF and TR response of both a WS\textsubscript{2} thin film and a single WS\textsubscript{2} screw-dislocation nanostructure.

\subsubsection{Transient-THG of a MoS\textsubscript{2} thin film vs. nanostructures}

\begin{figure}[!htbp]
	\centering
	\includegraphics[width=\linewidth]{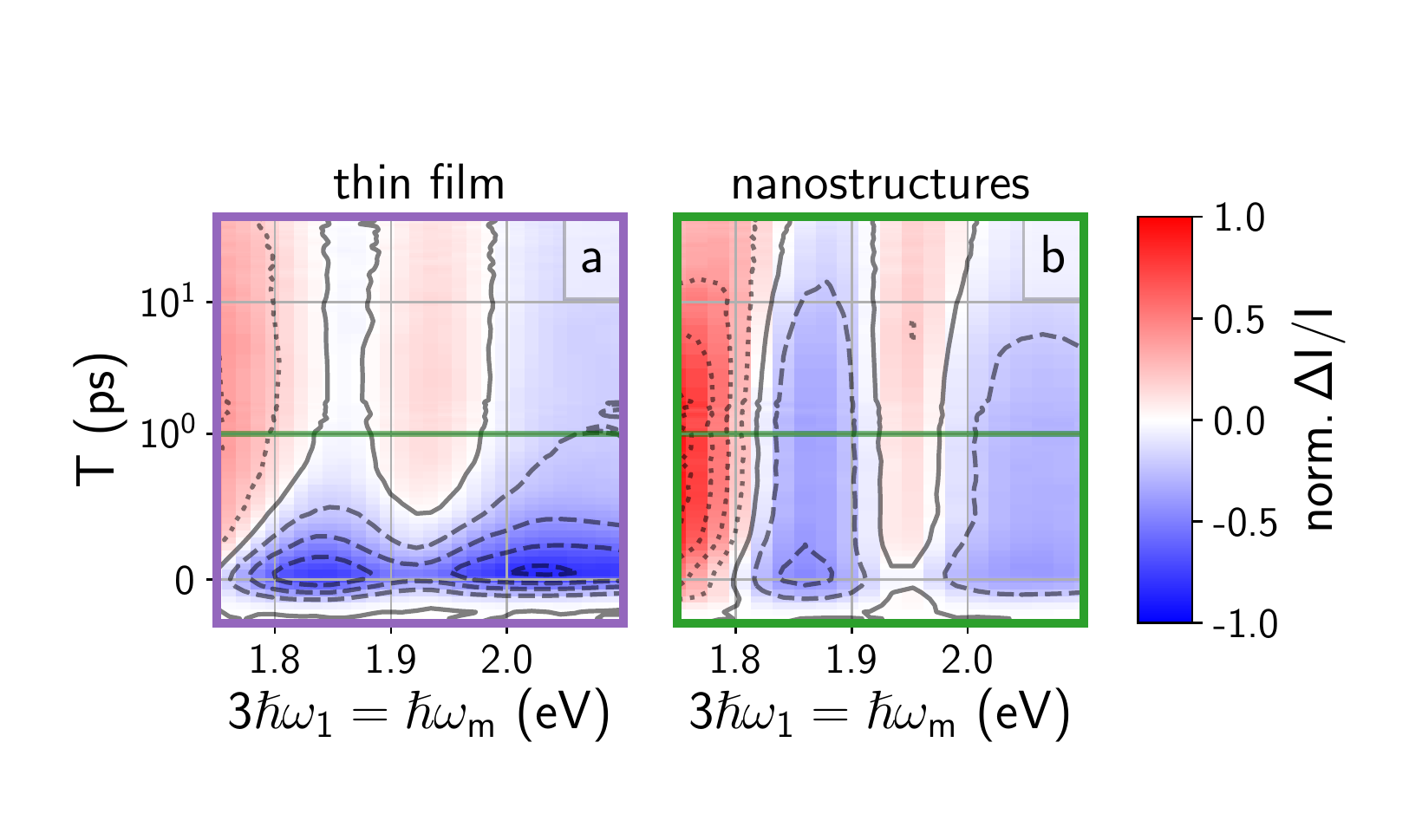}
	\caption{Pump-TSF-probe spectra of a MoS\textsubscript{2} thin film (a) and a MoS\textsubscript{2} spiral nanostructure ensemble (b). The temporal axis has linear scaling until 1 ps (green line) and then logarithmic scaling until the end (50 ps). In both frames $\hbar\omega_{\text{pump}}=3.1\text{ eV}$ with a fluence of 120 $\mu$J/cm\textsuperscript{2}.}
	\label{fig:morph}
\end{figure}

\autoref{fig:morph} shows the probe frequency vs. delay time response of the thin film (\autoref{fig:morph}a) and nanostructure (\autoref{fig:morph}b). 
Both spectra show similar spectral lineshapes near zero delay with decreases at the A and B features.
The nanostructures spectrum (\autoref{fig:morph}b) demonstrate narrower peaks and greater increase in TSF intensity to the red of the A exciton resonance compared to the thin film (\autoref{fig:morph}a).
The nanostructures exhibiting narrower features indicates that the thin film has more structural inhomogeneity.
While both samples show similar lineshapes at $T=0$, they exhibit drastically different dynamics.

\begin{figure}[!htbp]
	\centering
	\includegraphics[width=\linewidth]{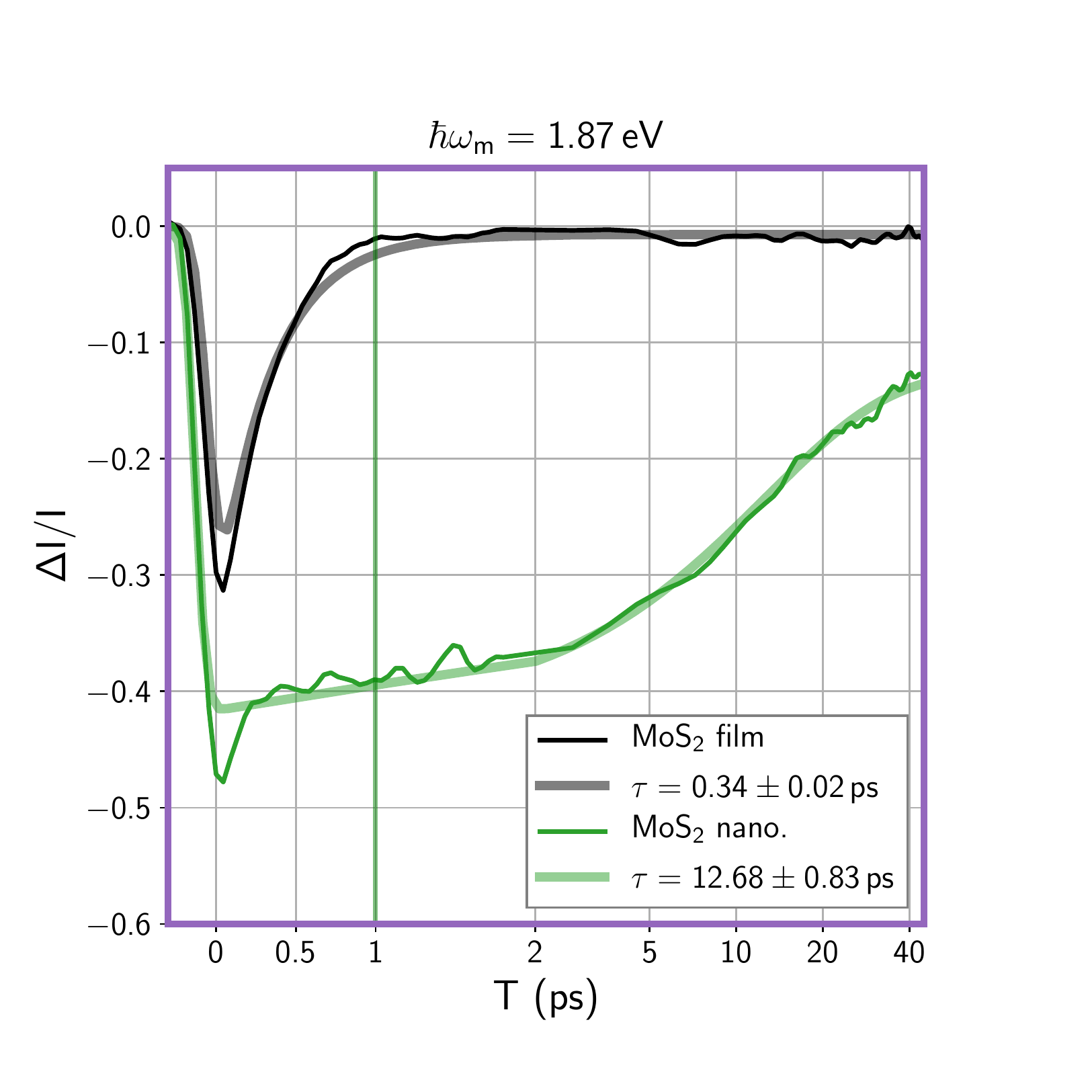}
	\caption{Pump-TSF-probe of a MoS\textsubscript{2} thin film  and an ensemble of spiral nanostructures. The temporal axis has linear scaling until 1 ps (green line) and then logarithmic scaling until the end (50 ps). This figure displays 1D slices out of \autoref{fig:morph} in which the pump is set to $\hbar\omega_{\text{pump}}=3.1\text{ eV}$ while the probe is set to $3\hbar\omega_1 = \hbar\omega_{m}=1.87\text{ eV}$ .}
	\label{fig:morph_dynamics}
\end{figure}

\autoref{fig:morph_dynamics} shows a single temporal trace through the data shown in \autoref{fig:morph} for each sample morphology. 
The thin lines are the measured data and the thick lines are fits using \autoref{E:fit}.
We recover exponential decay time constants of 0.34 $\pm$ 0.02 ps for the thin film and 12.7 $\pm$ 0.8 ps for the nanostructures. 
The morphology strongly affects dynamics.
It is likely the case that the dramatic differences in timescales are related to the density of grain boundaries, which can affect carrier scattering, recombination, and/or trapping.
The grain size of the thin film is on the order of 100 nm\textsuperscript{2} while that of the nanostructures is on the order of 10-100 $\mu$m\textsuperscript{2}.
We believe that carrier trapping is the main source of dynamics in \autoref{fig:morph_dynamics}; a carrier once it has been trapped is not able to efficiently screen the electron-hole Coulomb interaction, so BGR is lessened and the observed differential response is decreased.

There is a low amplitude, rapid dynamic present for each sample in \autoref{fig:morph_dynamics} that is not captured by our single exponential fit; we attribute this rapid dynamic to hot carriers (the carriers have $\sim$ 1 eV excess energy upon photoexcitation) cooling to form excitons.\cite{Ceballos_Zhao_2016, Cunningham_Jonker_2017}
In TMDCs, hot carriers bleach excitonic transitions more effectively than excitons; so a hot carrier cooling will reduce the bleach observed at the A and B transitions.\cite{SchmittRink_Miller_1985, Ceballos_Zhao_2016, Cunningham_Jonker_2017}
Taken together, we believe defect/grain-boundary assisted quenching of carriers along with hot carrier cooling can explain the dynamics observed in \autoref{fig:morph} and \autoref{fig:morph_dynamics}.

\subsubsection{Transient-THG vs. transient-reflectance for  WS\textsubscript{2} thin film vs. single nanostructure}

To further investigate the abilities of pump-TSF-probe, we performed pump-TSF-probe and pump-reflectance-probe experiments on two different morphologies of WS\textsubscript{2}: a thin film on a fused silica substrate (photo shown in \autoref{fig:characterization}a), and a single, isolated, spiral nanostructure grown on a 300 nm SiO\textsubscript{2} on Si substrate (optical microscope and atomic force microscope characterization shown in \autoref{fig:characterization}c, and \autoref{fig:characterization}b, respectively).
Our probe beam area is small compared to the area of the thin film, but much larger than the single nanostructure.

\begin{figure}[!htbp]
	\centering
	\includegraphics[width=\linewidth]{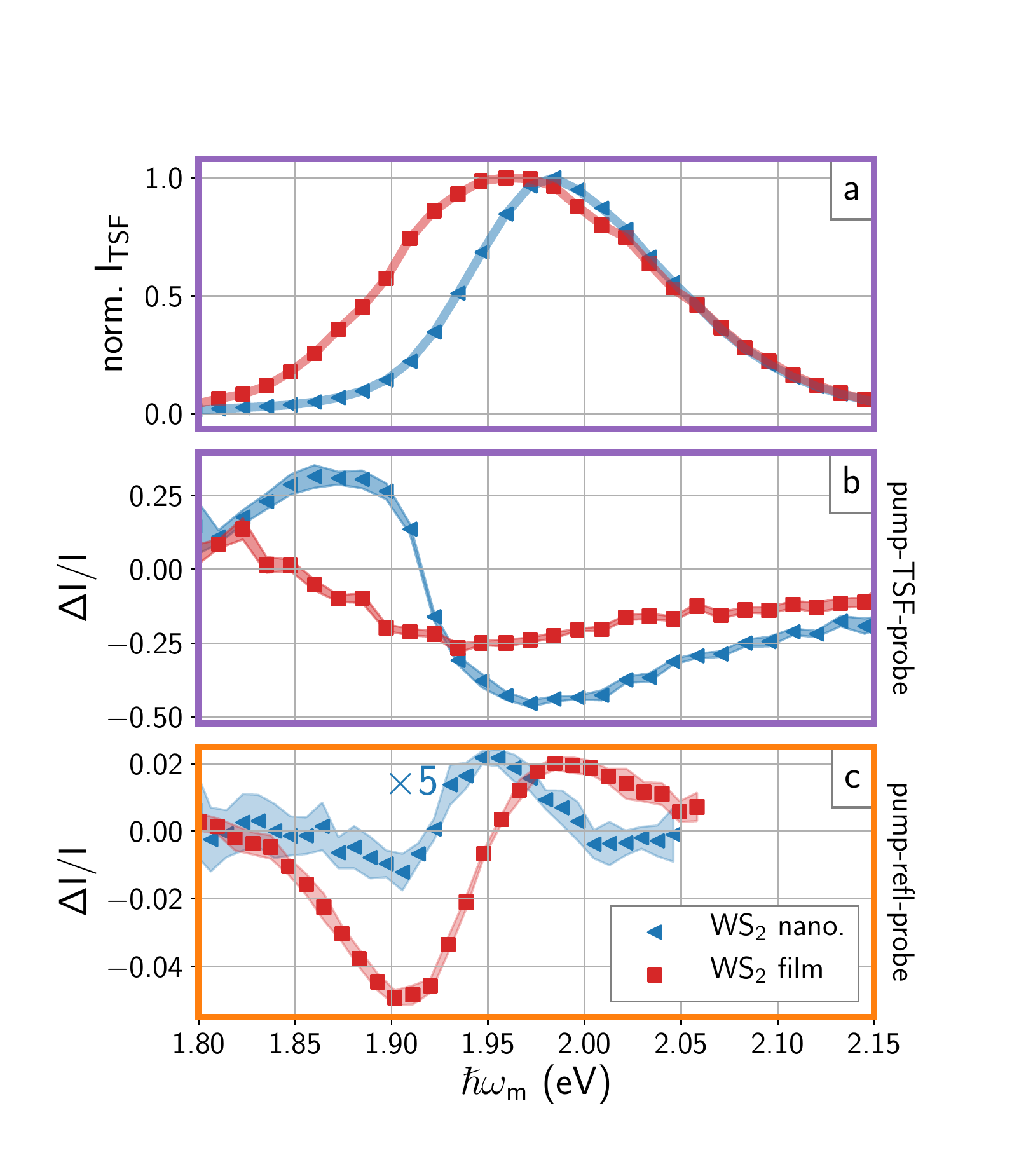}
	\caption{
		Comparison of pump-TSF-probe and pump-reflectance-probe for two morphologies of WS\textsubscript{2}: a thin film and a single, $\sim$150 $\mu$m wide spiral nanostructure. 
		(a) normalized TSF spectrum for each sample, these spectra are not normalized for the frequency dependent intensity of the probe OPA.
		(b) pump-TSF-probe spectra for each sample.
		(c) pump-refl-probe spectra for each sample. 
		In (b) and (c) the non-resonant pump has frequency of $\hbar\omega_{\text{pump}}=1.054\text{ eV}$ and a fluence of  $\sim$7000 $\mu$J/cm\textsuperscript{2}.
		All spectra were acquired for the same number of laser shots. 
		Each spectra is composed of 7 spectra averaged together at $T\approx0.12\text{ ps}$.
		(b) and (c) each show the standard deviation of the averaged spectra for each sample morphology as a filled spread around the average.}
	\label{fig:WS2_morph_spectra}
\end{figure}

In \autoref{fig:WS2_morph_spectra} we use a NIR pump to drive mid-gap or 2-photon transitions and probe the A exciton transition of WS\textsubscript{2}.
Appendix \ref{A:TR_NIR} describes experiments on our MoS\textsubscript{2} thin film which demonstrate our ability to drive mid-gap transitions with a NIR pump.
The full spectra and a discussion of these NIR pump measurements will be the subject of another publication. 
The unpumped THG spectra of the thin film and nanostructure are shown in \autoref{fig:WS2_morph_spectra}a, and the differential spectra ($T=120$ fs) for each morphology are shown in \autoref{fig:WS2_morph_spectra}b.
In both cases, the thin film exhibits a broader and redder A feature than the nanostructure---structural inhomogeneity from the small grain size of the polycrystalline film likely causes the increased linewidth of the thin film. 
The differing amount of spectral inhomogeneity causes the transient-reflectance and transient-TSF spectra between the two samples to look slightly different.

While we are able to measure clean transient-TSF spectra from both the thin film and single nanostructure, the same is less true for transient-reflectance measurements. 
\autoref{fig:WS2_morph_spectra}c shows that in addition to the qualitative differences in lineshape compared to the film, the nanostructure transient-reflectance signal is barely resolvable above measurement noise---it is at least 5 times weaker than the film's response.
Comparing the noise levels (width of the lines) between the two methods (\autoref{fig:WS2_morph_spectra}b and c) shows that pump-TSF-probe maintains a much higher signal-to-noise ratio than pump-reflectance-probe.
This sensitivity is due to the stronger scaling of TSF to transition dipole (\autoref{E:chimuJ}, $\mu^8$) compared to reflectance ($\mu^2$).
Since TMDC excitons interact strongly with light, the TSF emission from the substrates (fused silica and silicon) is negligible compared to the direct emission from the nanostructure.
In contrast, reflectance measurements are heavily dependent on the substrate and its refractive index (\autoref{E:R_thin_limit}).
As a result, TSF and pump-TSF-probe spectroscopies are insensitive to surface coverage and substrate layering, but reflectance and pump-reflectance-probe are sensitive to these effects.

\section{Outlook and Conclusion}\label{S:Conclusion}

This work shows that pump-TSF-probe spectroscopy can elucidate the dynamics and energetics of photoexcited semiconductor thin films and nanostructures using the examples of MoS\textsubscript{2} and WS\textsubscript{2}.
We demonstrated that this new spectroscopy (specifically the degenerate case of pump-TSF-probe, pump-THG-probe) and a more mature spectroscopy, transient-reflectance, can be understood in tandem from the same underlying physics.
We found that transient-TSF is robust to extrinsic nanocrystal properties, such as size and surface coverage, that dramatically affect transient-reflectance spectra.
This robustness allowed us to measure the transient-TSF spectrum of nanostructures much smaller than the excitation spot size, while at the same time maintaining a high signal-to-noise ratio.
We predict that with pump-TSF-probe spectroscopy, researchers can avoid the reflectance artifacts which have complicated recent ultrafast work (cf. refs \cite{Liu_Jin_2019, Ghosh_Ruhman_2018}) because the measured TSF and pump-TSF-probe intensities are easily connected to the samples susceptibility.

Previous studies have shown that TSF can be used to measure coupling between states.\cite{Handali_Wright_2018_B, Grechko_Bonn_2018}
Isolating cross peaks is a strategy not explored in this work that could further increase the selectivity of pump-TSF-probe spectroscopy in the future.
We believe that samples with states/bands of the correct symmetry would yield insightful dynamical coupling information. 
For instance, since TSF can measure the energy separations of up to four states, it could resolve how bands change their dispersion upon photo-excitation.

Another area that could benefit from the proof-of-concept measurements in this work is multi-photon microscopy.
Multiphoton microscopy uses a diverse set of techniques, including THG/TSF,\cite{Hanninen_Potma_2018_A, Hanninen_Potma_2018_B, Segawa_Hamaguchi_2012} to predominantly measure biologically relevant samples.
These multiphoton microscopies could easily incorporate a pump and a delay stage in order to measure spatially resolved dynamics. 

\section*{Supplementary Material}
	All data and the workup/representation/simulation scripts used to generate the figures in this work are available for download at \url{http://dx.doi.org/10.17605/OSF.IO/UMSXC}. 
	
\begin{acknowledgments}
	We acknowledge support from the Department of Energy, Office of Basic Energy Sciences, Division of Materials Sciences and Engineering, under award DE-FG02-09ER46664.	
	D.J.M. acknowledges support from the Link Foundation.
	We thank Kyle Czech for synthesizing the MoS\textsubscript{2} thin film sample.	
	D.J.M., D.D.K., and J.C.W. have filed a patent application on some of the work described herein.
\end{acknowledgments}

\appendix

\onecolumngrid

\hrulefill

\section{Calculation of single resonance pump-THG-probe response}\label{A:pthgp}

The single resonance third order susceptibility is given by
\begin{align}
\chi^{(3)} =  \frac{\mu^4}{\omega_{ag} - \omega_{321} - i\Gamma} \label{E:A1}.
\end{align}
We desire to calculate each term present in the total derivative
\begin{align}
\textrm{d}\chi^{(3)} &= \frac{\partial\chi^{(3)}}{\partial \mu}\textrm{d}\mu
+ \frac{\partial\chi^{(3)}}{\partial \omega_{ag}}\textrm{d}\omega_{ag}
+ \frac{\partial\chi^{(3)}}{\partial \Gamma}\textrm{d}\Gamma \label{E:A2}. 
\end{align}
By taking derivatives of \autoref{E:A1} we find
\begin{align}
\frac{\partial\chi^{(3)}}{\partial \mu}\textrm{d}\mu &= \frac{4\mu^3\textrm{d}\mu}{\omega_{ag} - \omega_{321} - i\Gamma} \\
\frac{\partial\chi^{(3)}}{\partial \omega_{ag}}\textrm{d}\omega_{ag} &= - \frac{\mu^4\textrm{d}\omega_{ag}}{\left(\omega_{ag} - \omega_{321} - i\Gamma\right)^2}\\
\frac{\partial\chi^{(3)}}{\partial \Gamma}\textrm{d}\Gamma &= \frac{i\mu^4\textrm{d}\omega_{ag}}{\left(\omega_{ag} - \omega_{321} - i\Gamma\right)^2}.
\end{align}
We now desire to calculate $\frac{\Delta I}{I}$
\begin{align}
\frac{\Delta I}{I} &= \frac{ \left|\chi^{(3)} + \textrm{d}\chi^{(3)}\right|^2 -  \left|\chi^{(3)}\right|^2}{ \left|\chi^{(3)}\right|^2} \\
&= \left|1 + \frac{\textrm{d}\chi^{(3)}}{\chi^{(3)}}\right|^2 - 1 \label{E:A8}
\end{align}
in which we have used the relationship $\frac{|a|}{|b|} = \left|\frac{a}{b}\right|$ for $b\neq0$.
Full substitution of \autoref{E:A1} and \autoref{E:A2} into \autoref{E:A8} yields a large equation which is too complicated to parse. 
A much simpler approach is to consider the limits of having only one of $\left\{  \textrm{d}\mu,\; \textrm{d}\omega_{ag},\; \textrm{d}\Gamma \right\}$ being nonzero at a time
\begin{align}
\frac{\Delta I}{I} &= 8\frac{\textrm{d}\mu}{\mu} + 16\left(\frac{\textrm{d}\mu}{\mu}\right)^2  && \textrm{d}\mu\neq 0\\
\frac{\Delta I}{I} &= \frac{\textrm{d}\Gamma}{\left(\omega_{ag}-\omega_{321}\right)^2+\Gamma^2} \left[-2\Gamma + \frac{\textrm{d}\Gamma}{\left(\omega_{ag}-\omega_{321}\right)^2+\Gamma^2}\left(\omega_{ag}^2 + \omega_{321}^2 + \Gamma^2 -2\omega_{ag}\omega_{321}\Gamma\right)\right]  && \textrm{d}\Gamma\neq 0\\
\frac{\Delta I}{I}  &= \frac{\textrm{d}\omega_{ag}}{\left(\omega_{ag}-\omega_{321}\right)^2+\Gamma^2} \left[-2\omega_{ag} + 2\omega_{321} + \frac{\textrm{d}\omega_{ag}}{\left(\omega_{ag}-\omega_{321}\right)^2+\Gamma^2}\left(\omega_{ag}^2 + \omega_{321}^2 + \Gamma^2 -2\omega_{ag}\omega_{321}\right)\right]  && \textrm{d}\omega_{ag}\neq 0 .
\end{align}
In the limit of small perturbation we may consider merely terms which are linear in all differentials
\begin{align}
\frac{\Delta I}{I} &\approx 8\frac{\textrm{d}\mu}{\mu}  && \textrm{d}\mu \neq 0\\
\frac{\Delta I}{I} &\approx \frac{-2\Gamma\textrm{d}\Gamma}{\left(\omega_{ag}-\omega_{321}\right)^2+\Gamma^2}   && \textrm{d}\Gamma \neq 0\\
\frac{\Delta I}{I}  &\approx \frac{2\textrm{d}\omega_{ag} \left(\omega_{321} -\omega_{ag}\right)}{\left(\omega_{ag}-\omega_{321}\right)^2+\Gamma^2}  && \textrm{d}\omega_{ag} \neq 0 .
\end{align}
This is the desired result which was discussed in the main text.
The lineshape for $\textrm{d}\mu\neq 0$ corresponds to a uniform change in the spectrum.
The lineshape for $\textrm{d}\Gamma\neq 0$ corresponds to the imaginary component of the original Lorentzian lineshape.
The lineshape for $\textrm{d}\omega_{ag}\neq 0$ corresponds to the first derivative lineshape of the original Lorentzian.

\section{Calculation of single resonance transient-transmittance response}\label{A:TT}

We desire to calculate the transient-transmittance response expected for a sample with a single resonance such that
\begin{align}
\chi^{(1)} =  \frac{\mu^2}{\omega_{ag} - \omega_{1} - i\Gamma}.
\end{align}
If we assume our samples are thick enough for Beer's law to apply, then the total amount of light with original intensity of $I_0$ transmitted through a sample of length $\ell$ is given by
\begin{align}
T = I_0 \exp{\left(-\alpha\ell\right)}
\end{align}
with $\alpha = \frac{2\pi  \text{Im}\left[ \chi^{(1)}\right]}{\lambda_1 n}=\frac{\omega_1  \text{Im}\left[ \chi^{(1)}\right]}{cn} $.
The transient-transmittance (absorbance) response can be constructed as
\begin{align}
\frac{\Delta T}{T} &= \frac{T_{\text{pumped}}-T_{\text{unpumped}}}{T_{\text{unpumped}}}\\
 &= \frac{I_0 \exp{\left(-\alpha_{\text{pumped}}\ell\right)}-I_0 \exp{\left(-\alpha_{\text{unpumped}}\ell\right)}}{I_0 \exp{\left(-\alpha_{\text{unpumped}}\ell\right)}} \\
&= \exp{\left(\alpha_{\text{unpumped}}\ell-\alpha_{\text{pumped}}\ell\right)} -1 .
\end{align}
Taylor expansion using $\exp{(x)} = 1 + x + \cdots$ yields
\begin{align}
\frac{\Delta T}{T} &\approx \ell \left(\alpha_{\text{unpumped}}-\alpha_{\text{pumped}}\right).
\end{align}
We now let $\alpha = \frac{\omega_1  \text{Im}\left[ \chi^{(1)}\right]}{cn} $ and 
$\chi^{(1)}_\text{pumped} = \chi^{(1)}_\text{unpumped} + \textrm{d}\chi^{(1)}$ which yields
\begin{align}
\frac{\Delta T}{T} &\approx \ell \left( \frac{\omega_1\text{Im}\left[ \chi^{(1)}_\text{unpumped} \right]}{cn} - \frac{\omega_1 \text{Im}\left[\chi^{(1)}_\text{unpumped} + \textrm{d}\chi^{(1)}\right]}{cn}\right) \\
&= -\frac{\omega_1\ell}{cn}\text{Im}\left[\textrm{d}\chi^{(1)}\right].
\end{align}
In the case of small perturbation, $\textrm{d}\chi^{(1)}$ may be described by the total derivative 
\begin{align}
\textrm{d}\chi^{(1)} &= \frac{\partial\chi^{(1)}}{\partial \mu}\textrm{d}\mu
+ \frac{\partial\chi^{(1)}}{\partial \omega_{ag}}\textrm{d}\omega_{ag}
+ \frac{\partial\chi^{(1)}}{\partial \Gamma}\textrm{d}\Gamma . \label{E:dchi1}
\end{align}
Each type of change ($\textrm{d}\mu$, $\textrm{d}\omega_{ag}$, $\textrm{d}\Gamma$) affects the observed lineshape of $\frac{\Delta T}{T}$.
We consider the cases where only one of the differentials is non-zero at a time.
The partial derivatives and imaginary projects are trivial and the results are
\begin{align}
\frac{\Delta T}{T} &= -\left(\frac{\omega_1\ell}{cn}\right)\frac{2\mu \Gamma\textrm{d}\mu}{\left(\omega_{ag}-\omega_{1}\right)^2 + \Gamma^2}   && \textrm{d}\mu\neq 0\\
\frac{\Delta T}{T} &= -\left(\frac{\omega_1\ell}{cn}\right)\frac{2\mu^2 \left(\left(\omega_{ag}-\omega_{1}\right)^2-\Gamma^2\right)\textrm{d}\Gamma}{\left(\left(\omega_{ag}-\omega_{1}\right)^2 + \Gamma^2\right)^2}   && \textrm{d}\Gamma\neq 0\\
\frac{\Delta T}{T}  &= \left(\frac{\omega_1\ell}{cn}\right)\frac{2\mu^2 \Gamma\left(\omega_{ag}-\omega_{1}\right)\textrm{d}\omega_{ag}}{\left(\left(\omega_{ag}-\omega_{1}\right)^2 + \Gamma^2\right)^2}  && \textrm{d}\omega_{ag}\neq 0 .
\end{align}
The lineshape for $\textrm{d}\mu\neq 0$ corresponds to the imaginary component of the original Lorentzian lineshape.
The lineshape for $\textrm{d}\Gamma\neq 0$ corresponds to the second derivative lineshape of the original Lorentzian.
The lineshape for $\textrm{d}\omega_{ag}\neq 0$ corresponds to the first derivative lineshape of the original Lorentzian.

\section{Lineshape modeling}\label{A:lineshapes}

In this appendix we describe our simple model for building the spectral lineshapes shown in \autoref{fig:fitting}.
The general implementation is:
\begin{enumerate}
	\item For both spectroscopies construct an unexcited $\chi^{(n)}$ spectrum from a sum of oscillators.
	\item Calculate the unexcited reflectance or TSF spectrum from $\chi^{(1)}$ and $\chi^{(3)}$, respectively. 
	\item Create a $\chi^{(n)\prime}$ spectrum to perturb the central frequencies, linewidths, and amplitudes of the oscillators used to construct $\chi^{(n)}$.
	\item Calculate the excited reflectance or TSF spectrum from $\chi^{(1)\prime}$ and $\chi^{(3)\prime}$, respectively.  
	\item Use \autoref{E:dXX} to calculate $\frac{\Delta I}{I}$ for both spectroscopies.
	\item Iterate through previous steps to fit observed lineshapes.
\end{enumerate}
We choose to use complex, Lorentzian oscillators to construct our spectra:
\begin{align}
\chi^{(n)} = \sum_{j=1} \sqrt{\frac{\Gamma_j}{\pi}}\frac{A_j}{E_{0,j}-\hbar\omega_m - i\Gamma_j}
\end{align}
in which $j=1$ and $j=2$ are the A and B transitions, and the other oscillators are high-lying non-resonant transitions.
To create $\chi^{(n)\prime}$ we replace $\Gamma_j \rightarrow \Gamma_j + \Delta\Gamma_j$, $E_{0,j} \rightarrow E_{0,j} + \Delta E_{0,j}$, and $A_j \rightarrow (1-\%\textrm{ decrease})A_j$.
ESA-like additional transitions are incorporated by adding a phased offset to $\chi^{(n)\prime}$; the pump-TSF-probe spectrum in \autoref{fig:fitting}b has a slight offset with phase described by $\exp\left[i\theta\right]$ with $\theta=1\text{ radian}$. 
\autoref{T:fitparams} codifies the parameters we found, by hand, to give acceptable fits to the data shown in \autoref{fig:fitting}.

We construct a TSF spectrum by merely calculating the square magnitude of $\chi^{(3)}$ as indicated by \autoref{E:ITSF}.
We construct a reflectance spectrum by converting $\chi^{(1)}$ to a complex refractive index, $\bar{n}$ and then using a Fresnel-coefficient-like analysis, specifically as discussed in \textcite{Anders_1967}, which takes into account the influence of multiple reflections and the substrate.
This treatment is slightly more holistic than merely using \autoref{E:R} because it takes into account the finite thickness of the sample (while the derivation of \autoref{E:R} assumes a delta function sample).
$R$ is calculated using
\begin{align}
R &= \left| \frac{\bar{r}_1 + \bar{r}_2\exp{\left[-i\phi_1\right]}}{1+\bar{r}_1\bar{r}_2\exp{\left[-i\phi_1\right]}}\right|^2 \\
\bar{r}_1 &= \frac{\bar{n}_0-\bar{n}_1}{\bar{n}_0+\bar{n}_1} \\
\bar{r}_2 &= \frac{\bar{n}_1-\bar{n}_2}{\bar{n}_1+\bar{n}_2} \\
\phi_1 &= \frac{4\pi\ell\bar{n}_1}{\lambda}
\end{align}
in which $\bar{n}_0$ is the refractive index of air, $\bar{n}_1$ is the constructed refractive index of the MoS\textsubscript{2} thin film with thickness $\ell$, $\bar{n}_2$ is the refractive index of the fused silica substrate, and $\lambda$ is the vacuum wavelength of the interrogating electric field.

\begin{table*}[!htbp]
	\caption{\label{T:fitparams} Parameters used to produce the lineshapes shown in \autoref{fig:fitting}.Note that the model in \autoref{fig:fitting}b for pump-TSF-probe has a slight offset with phase described by $\exp\left[i\theta\right]$ with $\theta=1\text{ radian}$ and amplitude of 1\% of the maximum feature.}
	\begin{ruledtabular}
		\begin{tabular}{lllllll}
			transition & $E_0$ (eV) & $\Delta E_0$ (eV) & $\Gamma$ (eV) & $\Delta\Gamma$ (eV) & relative $A$     & \% $A$ decrease \\ \hline
			\multicolumn{7}{c}{\textbf{transient reflectance model $T=0.05$ ps}}             \\
			A & 1.807   & -0.01   & 0.1    & 0.002   & 1     & 2          \\
			B & 1.98    & -0.009  & 0.12   & 0.005   & 1.1   & 2          \\
			& 2.7     & -0.008        & 0.25   & 0       & 4     & 5          \\
			& 3.2     & 0        & 0.25   & 0       & 8     & 0          \\
			& 6       & 0        & 0.25   & 0       & 40    & 0          \\ \hline
			\multicolumn{7}{c}{\textbf{transient reflectance model $T=0.55$ ps}}             \\
			A & 1.807   & -0.005   & 0.1    & 0.002   & 1     & 2          \\
			B & 1.98    & -0.005  & 0.12   & 0.005   & 1.1   & 2          \\
			& 2.7     & 0        & 0.25   & 0       & 4     & 5          \\
			& 3.2     & 0        & 0.25   & 0       & 8     & 0          \\
			& 6       & 0        & 0.25   & 0       & 40    & 0          \\ \hline
			\multicolumn{7}{c}{\textbf{transient TSF model $T=0.05$ ps}}                    \\
			A & 1.81   & -0.012   & 0.085  & 0.005   & 1     & 2        \\
			B & 1.95    & -0.009  & 0.1    & 0.005   & 0.91  & 2          \\  \hline       
			\multicolumn{7}{c}{\textbf{transient TSF model $T=0.55$ ps}}                    \\
			A & 1.81   & -0.003   & 0.085  & 0  & 1     & 0        \\
			B & 1.95    & -0.002  & 0.1    & 0   & 0.91  & 0          \\   
		\end{tabular}
	\end{ruledtabular}
\end{table*}

\clearpage

\section{Transient-reflectance with NIR excitation of a MoS\textsubscript{2} thin film}\label{A:TR_NIR}

\begin{figure}[!htbp]
	\centering
	\includegraphics[width=0.5\linewidth]{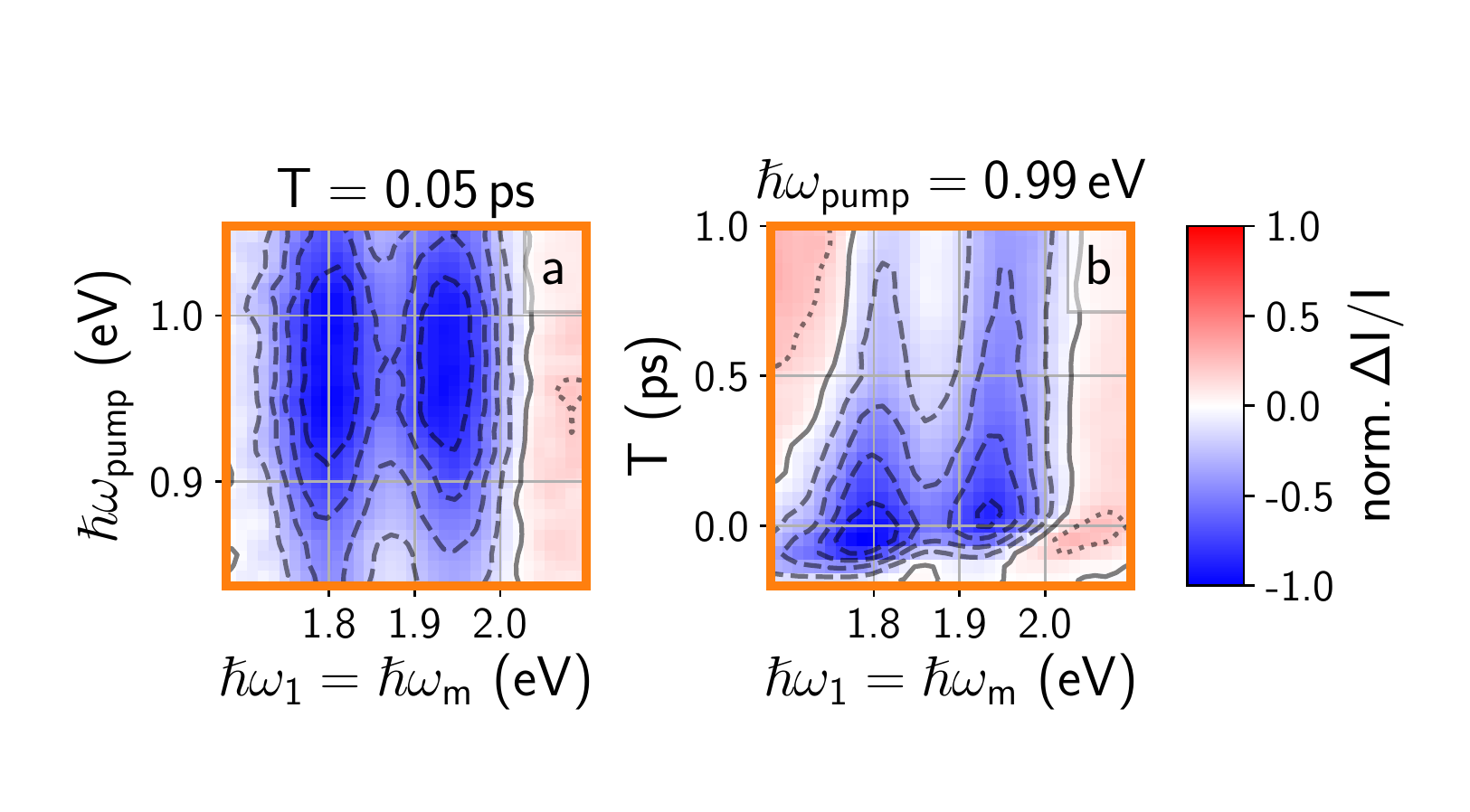}
	\caption{Transient-reflectance spectroscopy on a MoS\textsubscript{2} thin film with a NIR pump. (a) shows the transient-reflectance spectrum for different combinations of pump and probe frequencies for $T=50\text{ fs}$. Note that this spectrum is not normalized for the setpoint frequency dependence of the pump laser intensity. (b) shows the measured dynamics for different probe colors with $\hbar\omega_{\text{pump}} = 0.99\text{ eV}$}
	\label{fig:NIR_pump}
\end{figure}

TMCDs are known to be weakly absorptive well below bandgap (c.f. \textcite{Bikorimana_Kulyuk_2016}). 
To investigate this sub-band edge response, we tuned our pump to NIR colors, using fluence an order of magnitude higher than the visible pump. 
The effects of this sub-band edge pump on the band-edge reflectance spectrum are shown in \autoref{fig:NIR_pump}.
We observe similar spectral and temporal lineshapes for both the visible and NIR pump, indicating the NIR pump generates photocarriers in a similar manner to a visible pump.

\begin{figure}[!htbp]
	\centering
	\includegraphics[width=0.5\linewidth]{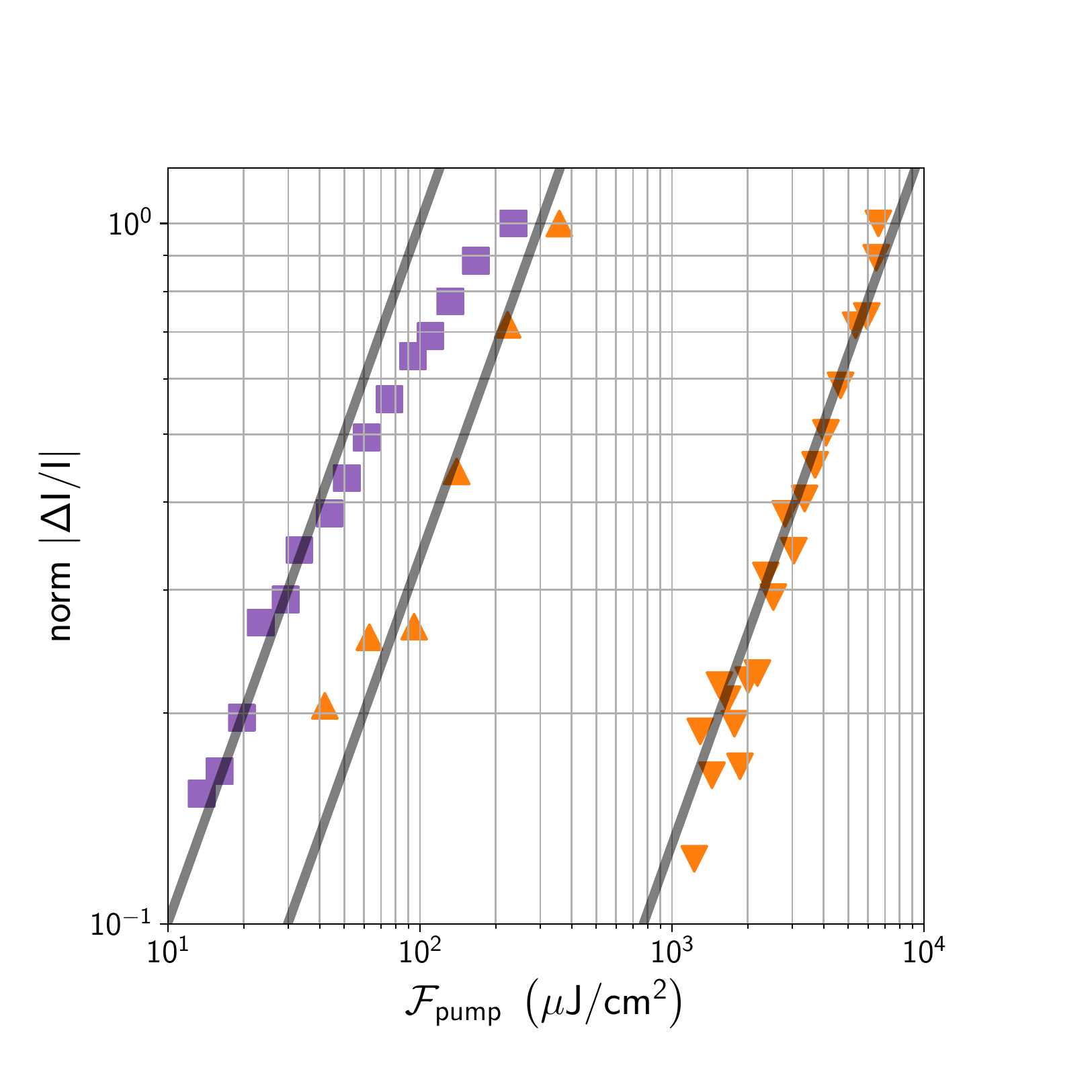}
	\caption{
		Comparison of transient-reflectance spectroscopy (orange) to transient-TSF spectroscopy (violet) pump fluence scaling for a MoS\textsubscript{2} thin film.
		The y-axis is maximum extent of the bleach measured (near $T=0$). The pump and probe combinations are as follows: ($\blacktriangle$, $\hbar\omega_1 = \hbar\omega_m =\hbar\omega_{\text{pump}}=1.98\text{ eV}$);  ($\blacktriangledown$, $\hbar\omega_1 = \hbar\omega_m =1.98\text{ eV}$, $\hbar\omega_{\text{pump}} =0.99\text{ eV}$); and ($\blacksquare$, $3\hbar\omega_1 = \hbar\omega_m =2.05\text{ eV}$, $\hbar\omega_{\text{pump}} =1.98\text{ eV}$).
		Gray lines are guides to the eye signifying linear scaling of response with pump fluence.}
	\label{fig:fluence}
\end{figure}

Given the strong two-photon absorption in TMDCS,\cite{Zhang_Wang_2015, Ye_Zhang_2014, Berkelbach_Reichman_2015, Dong_Wang_2018, Cui_Xu_2018} it is reasonable to attribute the signals in \autoref{fig:NIR_pump} to two-photon absorption from the pump.
We find, however, that the TR and pump-THG-probe scale linearly or sublinearly, rather than quadratically, with pump fluence (\autoref{fig:fluence}).
Furthermore, pump-induced reflectance responses occur with pump photon energies below half the band edge. 
These observations rule out two-photon absorption as the dominant contribution to \autoref{fig:NIR_pump}.

We conclude that our NIR pump excites electrons/holes to/from midgap states which have small optical cross-sections.
Midgap states are known to exist in synthetically grown MoS\textsubscript{2} and are generally attributed to sulfur vacancies and edge defects.\cite{Cunningham_Hayden_2016, vanderZande_Hone_2013, Zhou_Idrobo_2013, Yu_Yakobson_2015, Hong_Zhang_2015, Qiu_Wang_2013, Lu_Andrei_2014}
We believe mid-gap excitations can induce BGR and band-filling in a manner similar to direct, allowed transitions, which explains the similarity between visible and NIR pumps (compare \autoref{fig:freq} a with \autoref{fig:NIR_pump}a or \autoref{fig:wigner}a with \autoref{fig:NIR_pump}b).
The insensitivity to pump wavelength reflects the large dispersion of mid-gap states and their transitions to valence and conduction bands.

%%% finish up %%%%%%%%%%%%%%%%%%%%%%%%%%%%%%%%%%%%%%%%%%%%%%%%%%%%%%%%%%%%%%%%%%%%%%%%%%%%%%%

\clearpage

\twocolumngrid

\bibliography{database}

\end{document}